\pdfoutput=1
\documentclass[aps,twocolumn,amsmath,amssymb,reprint,superscriptaddress,notitlepage]{revtex4-1}
\usepackage{graphicx}
\usepackage{latexsym}
\usepackage{times}
\usepackage{color}
\usepackage{amsmath}
\usepackage{dsfont}
\usepackage{dcolumn}
\usepackage{multirow}
\usepackage{mathbbol}
\usepackage{balance}
\usepackage{latexsym,amsmath,amssymb,bm,euscript}
\usepackage{xr}
\usepackage{lipsum}
\usepackage[normalem]{ulem}

\usepackage{esvect}
\usepackage{color} 

\usepackage{hyperref}
\hypersetup{
colorlinks=true,final=true,
        linkcolor=blue,
        citecolor=blue,
        filecolor=blue,
        urlcolor=blue
}

\newcommand{\ket}[1]{| #1 \rangle}

\renewcommand{\H}[0]{H}
\newcommand{\hc}{^{\dagger}}							

\newcommand{\nn}{\nonumber}
\renewcommand{\l}[0]{\left}
\renewcommand{\r}[0]{\right}	
\newcommand{\ii}{\mathrm{i}}
\newcommand{\cc}{^{\ast}}
\newcommand{\comm}[2]{\left[ #1, #2 \right]}
\newcommand{\diss}[1]{\mathcal{D}[ #1 ]} 	  	
\newcommand{\Tr}{\text{Tr}}
\newcolumntype{d}[1]{D{.}{.}{#1}}

\begin{document}

\title{Microwave photon-mediated interactions between semiconductor qubits}
\author{D.~J.~van~Woerkom}
\thanks{These authors contributed equally to this work.}
\affiliation{Department of Physics, ETH Zurich, CH-8093 Zurich, Switzerland}

\author{P.~Scarlino}
\thanks{These authors contributed equally to this work.}
\affiliation{Department of Physics, ETH Zurich, CH-8093 Zurich, Switzerland}

\author{J.~H.~Ungerer}
\affiliation{Department of Physics, ETH Zurich, CH-8093 Zurich, Switzerland}

\author{C.~M\"uller}
\affiliation{Department of Physics, ETH Zurich, CH-8093 Zurich, Switzerland}

\author{J.~V.~Koski}
\affiliation{Department of Physics, ETH Zurich, CH-8093 Zurich, Switzerland}

\author{A.~J.~Landig}
\affiliation{Department of Physics, ETH Zurich, CH-8093 Zurich, Switzerland}

\author{C.~Reichl}
\affiliation{Department of Physics, ETH Zurich, CH-8093 Zurich, Switzerland}

\author{W.~Wegscheider}
\affiliation{Department of Physics, ETH Zurich, CH-8093 Zurich, Switzerland}

\author{T.~Ihn}
\affiliation{Department of Physics, ETH Zurich, CH-8093 Zurich, Switzerland}

\author{K.~Ensslin}
\affiliation{Department of Physics, ETH Zurich, CH-8093 Zurich, Switzerland}

\author{A.~Wallraff}
\affiliation{Department of Physics, ETH Zurich, CH-8093 Zurich, Switzerland}

\date{\today}

\begin{abstract}
The realization of a coherent interface between distant charge or spin qubits in semiconductor quantum dots is an open challenge for quantum information processing. Here we demonstrate both resonant and non-resonant photon-mediated coherent interactions between double quantum dot charge qubits separated by several tens of micrometers. We present clear spectroscopic evidence of the collective enhancement of the resonant coupling of two qubits. With both qubits detuned from the resonator we observe exchange coupling between the qubits mediated by virtual photons. In both instances pronounced bright and dark states governed by the symmetry of the qubit-field interaction are found. Our observations are in excellent quantitative agreement with master-equation simulations.
The extracted two-qubit coupling strengths significantly exceed the linewidths of the combined resonator-qubit system.
This indicates that this approach is viable for creating photon-mediated two-qubit gates in quantum dot based systems.
\end{abstract}
\maketitle




Semiconductor nanostructure based systems are one of the promising contenders for quantum information processing since they offer flexibility in tuning, long coherence times and well-established fabrication techniques~\cite{Hanson2007,Awschalom2013}. However, scaling to larger numbers of qubits remains a challenge, since many coupling mechanisms for realizing two-qubit gates are short range, i.e.~limited to nearest neighbors. For scaling to larger systems and eventually to a full scale quantum computer, a combination of short and longer range interactions seems promising~\cite{Vandersypen2017}.

So far, short range ($\sim$~100~nm) qubit-qubit interaction has been realized via capacitive or exchange coupling between charge~\cite{shinkai2009,H.O.li2015,ward2016} and spin qubits~\cite{Shulman2012,Veldhorst2015,Watson2017a,Zajac2017}, 
which was expanded by making use of interactions mediated by additional qubits ($\sim\,400$~nm)~\cite{Baart2016c} or electronic cavities ($\sim 1.7 \, \rm{\mu}$m) \cite{Nicoli2017}. 
However, it is predicted that the range of interaction between semiconductor qubits can be increased significantly using microwave photons~\cite{Childress2004,Burkard2006,Vandersypen2017}. 
A key ingredient, the strong coupling of individual charges \cite{Mi2017,Stockklauser2017} 
or spins \cite{Mi2017d,Samkharadze2017,Landig2017} to individual microwave photons, has recently been realized in semiconductor implementations of circuit quantum electrodynamics (QED) \cite{Wallraff2004}.

Here, we present experiments in which the coherent photon-mediated coupling between two spatially separated semiconductor qubits is realized both in the resonant and the dispersive regime using high impedance SQUID array resonators. The high Josephson inductance of the SQUID array increases the strength of the vacuum fluctuations of the electric field, enhancing the coupling strength of the individual qubits to the resonator \cite{Stockklauser2017} and consequently the qubit-qubit coupling, which allows us to overcome the limitations of prior experiments~\cite{Delbecq2013,Deng2015d,Mi2017d}.
This key step holds the strong promise that two-qubit gates based on photon-mediated interactions, which are a corner-stone in quantum information processing with superconducting circuits~\cite{Otterbach2017}
, are implementable with semiconductor qubits based on a variety of material systems.







\begin{figure}[b!]
\includegraphics[width=0.50\textwidth]{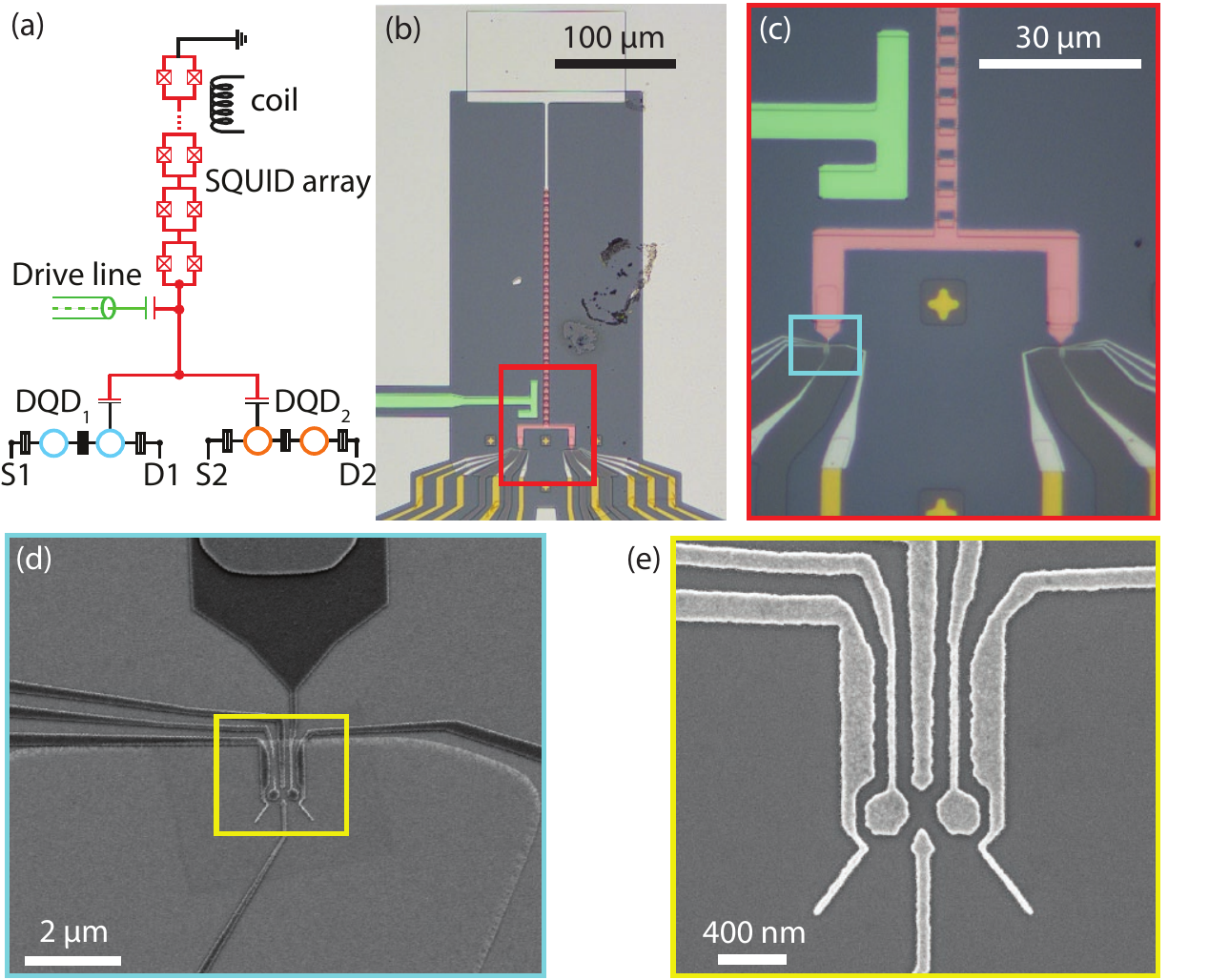}
\caption{Simplified circuit diagram and micrograph of the device. (a) Schematic of the device and control line: SQUID array resonator (red), drive line (green), two DQDs (cyan and orange) and an external coil (black). Color code is used throughout the manuscript. (b) False color optical micrograph of the measured device. (c) Detail of (b) showing the resonator and its drive line coupled to both DQDs. (d) Scanning electron micrograph (SEM) of the resonator connected to DQD$_1$ on the left. DQD$_2$ is defined as a mirrored copy of DQD$_1$, separated from it by 42~$\mu$m. (e) SEM micrograph of gate structure used for defining the DQDs in the GaAs/AlGaAs heterostructures. (d) and (e) are images of identically designed devices not used in the experiments.}
\label{fig1:SampleAndCircuit}
\end{figure}

In this work, we investigate two semiconductor double quantum dot (DQD) charge qubits strongly coupled to a single high-impedance resonator [Figs.~\ref{fig1:SampleAndCircuit}(a) and (b)] composed of 35~SQUIDS with an estimated impedance of $\sim$~1~k$\Omega$~\cite{Stockklauser2017}.
At the end of the flux tunable resonator, the DQDs are defined using depletion gate technology on a mesa of a GaAs/AlGaAs heterostructure. They are separated by a distance of 42~$\mu$m, much larger than in conventional multi quantum dot devices. Both DQD charge qubits are coupled to the antinode of the electric field at the open end of the resonator [Figs.~\ref{fig1:SampleAndCircuit}(c)-(e)]. The resonator is designed for read-out and additionally acts as a coupler between the spatially separated DQDs. The design and fabrication is similar to the one described in Refs.~\cite{Stockklauser2017,Scarlino2017b} and is discussed in detail in the Appendix~\ref{sec:fab}.


We characterize the properties of the device by measuring the amplitude $|S_{11}|$ and phase $\varphi$ of a microwave tone reflected off the resonator at the drive line indicated in green in Fig.~\ref{fig1:SampleAndCircuit}. The same line is also used to apply microwave spectroscopy tones to the individual qubits (see Appendix~\ref{sec:setup} for a complete description of the measurement setup).

With the qubit transition frequencies far detuned from the 
resonator~\cite{Stockklauser2017} operated at $\omega_\mathrm{r}/2\pi=5.171$~GHz, we spectroscopically determine the resonator internal loss rate $\kappa_\mathrm{int}/2\pi=17$~MHz, dominated by the residual coupling to the gate leads~\cite{Mi2017c}, and its external coupling rate $\kappa_\mathrm{ext}/2\pi=6$~MHz, governed by the coupling to the drive line. This puts the resonator into the weakly undercoupled regime ($\kappa_\mathrm{int}>\kappa_\mathrm{ext}$) keeping the total resonator line width small. We configure the two DQDs ($k=1,2$) as two-level systems described in good approximation by the Hamiltonian $H_k = - {\delta}_{k} \sigma_{z}/2 + {t}_{k} \sigma_{x}$ with Pauli matrices $\sigma_{x,y,z}$. The transition frequency of each quantum dot qubit  $\omega_{\text{DQD-}k}=\sqrt{4t_k^2+\delta_k^2}$ is a hyperbolic function of detuning ${\delta}_k$ between the charge states of the individual dots and tunnel rate $2{t}_k$ between them.

\begin{figure}[ht!]
\includegraphics[width=0.50\textwidth]{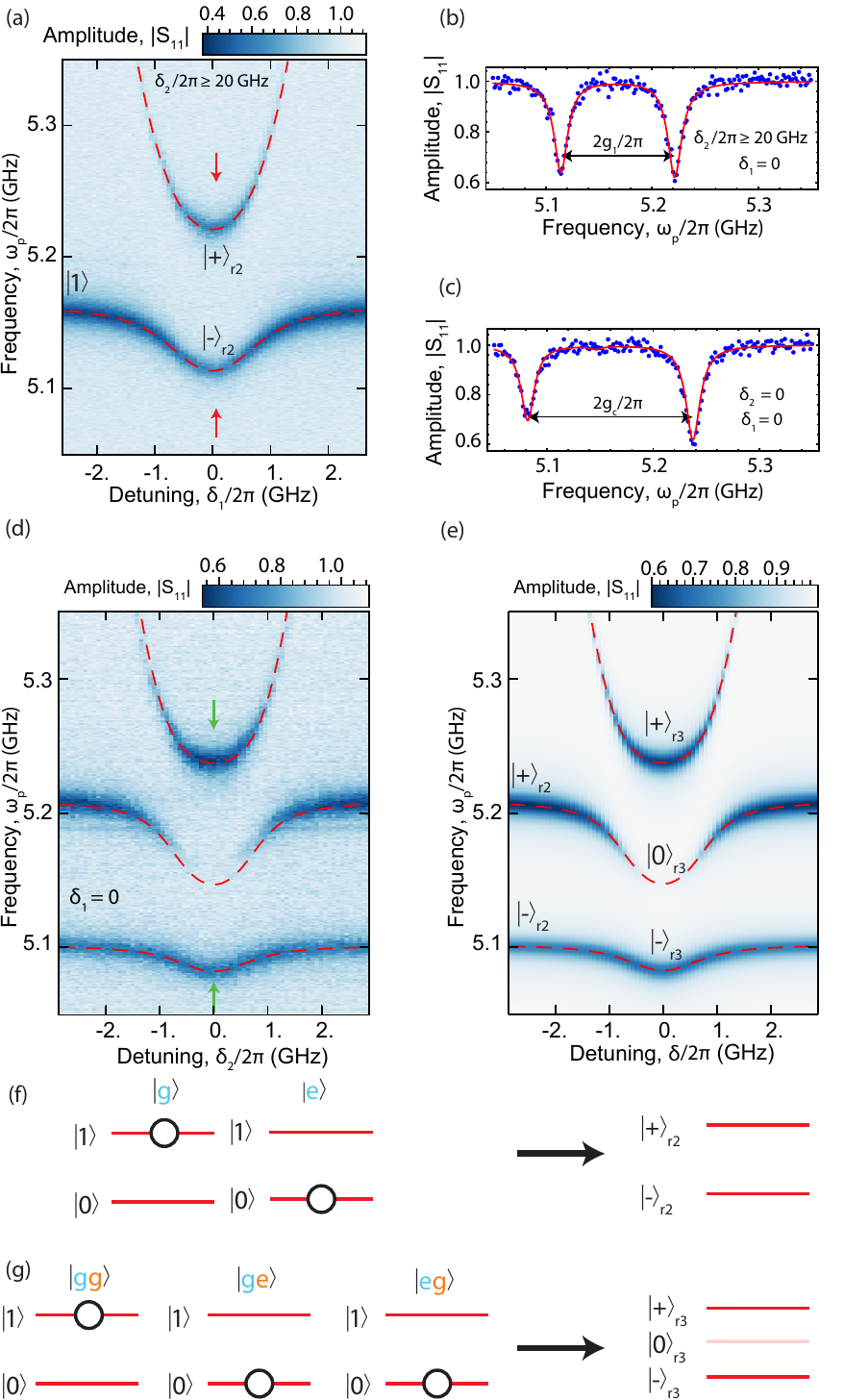}
\caption{Individual and two-qubit vacuum Rabi mode splitting. (a) Measured reflection coefficient $|S_{11}|$ \emph{vs}.~drive frequency $\omega_\mathrm{p}$ and charge detuning $\delta_1$ for DQD$_1$ ($2t_1/2\pi=5.166$~GHz) tuned into resonance with the SQUID array resonator ($\omega_\mathrm{r}/2\pi=5.171$~GHz). The red-dashed lines are extracted from fits to a master equation model, see text for details. (b) $|S_{11}|(\omega_\mathrm{p})$ at $\delta_1=0$ [red arrows in (a)]. (c) $|S_{11}|(\omega_\mathrm{p})$ at $\delta_1 \sim \delta_2 \sim 0$ [green arrows in (d)]. (d) $|S_{11}|$ \emph{vs.}~drive frequency $\omega_\mathrm{p}$ and charge detuning $\delta_2$ for DQD$_1$ biased at $2t_1/2\pi=5.166$~GHz and $\delta_{1}\approx0$ and DQD$_2$ at $2t_2/2\pi=5.156$~GHz realizing $\omega_\mathrm{r}=\omega_\mathrm{DQD1}=\omega_\mathrm{DQD2}$ at $\delta_{1,2}\approx0$. (e) Master equation simulation of $|S_{11}|(\delta_2,\omega_\mathrm{p})$ fitting to the data in panel (d), details in Appendix~\ref{sec:Model_sys}.  Schematic energy level diagram of (f) one and (g) two DQDs interacting with the resonator. Basis states are shown on the left, coupled states on the right, see text and Appendix~\ref{sec:Model_sys} for state labeling.}
\label{fig2:OnResonance}
\end{figure}

We tune $2t_1 \sim \omega_\text{r}$ such that DQD$_1$ is in resonance with the high impedance resonator at $\delta_1 = 0$. We first measure the reflection spectrum $|S_{11}|$ of the resonator tuning $\delta_1$ of the left DQD, labeled DQD$_1$ in Fig.~\ref{fig1:SampleAndCircuit}(a), from negative to positive values with the second DQD far detuned, $\delta_2 \gg \omega_\text{r}$. We observe a well resolved vacuum Rabi mode splitting~\cite{Stockklauser2017} with a coupling rate of $g_1/2\pi$~=~53~MHz. The photon state of the resonator and the charge state of the DQD hybridize in a resonant two-body (anti)symmetric state, $|\pm\rangle_{\rm{r2}}=(|e,0\rangle \pm|g,1\rangle)/\sqrt{2}$~\cite{Blais2004} as illustrated in Fig. \ref{fig2:OnResonance}(f) with the charge qubit ground $|g\rangle$ and excited state $|e\rangle$, and the cavity photon number states $|0\rangle$, $|1\rangle$.
We independently determined the linewidth $\Gamma_{2,1}/2\pi=4.8\pm0.6$~MHz of DQD$_1$ at this frequency using qubit spectroscopy in the dispersive regime making use of the tunable resonator~\cite{Stockklauser2017}. Equivalent measurements were performed for DQD$_2$, adjusting its bias configuration to reach a coupling strength $g_2/2\pi=56$~MHz, similar to DQD$_1$, and finding $\Gamma_{2,2}/2\pi=5.6\pm0.5$~MHz, see Appendix~\ref{opposite_fig2}. These results shows that each DQD is individually strongly coupled to the resonator, $g_k>\left(\kappa_\text{int}+\kappa_\text{ext}\right)/2+\Gamma_{2,k}$.

We model the coupled system using a mater equation simulation with the Tavis-Cummings Hamiltonian
\begin{align}\label{eq:HTot:main}
	\H = \omega_\text{r} a\hc a + \sum_{k} H_k + \sum_{k} g_{k} \sigma_{z} (a\hc + a)\,,
\end{align}
with $\hbar=1$, and bosonic annihilation (creation) operators $a$ ($a\hc$), and the coupling rate $g_k$ between the resonator and DQD$_k$ (Appendix~\ref{sec:Model_sys}). The observed resonance frequencies and linewidths are in excellent agreement with the simulation (dashed lines in Fig.~\ref{fig2:OnResonance}) which allow us to extract the system parameters with high accuracy (Appendix~\ref{sec:para}).

We now explore the case of all three transitions tuned into mutual resonance, ($\omega_\mathrm{r}=\omega_\mathrm{DQD1}=\omega_\mathrm{DQD2}$) by measuring the reflection spectrum $|S_{11}|$ of DQD$_1$ resonantly coupled to the resonator ($2t_1=\omega_\mathrm{r}$, $\delta_1=0$) and tuning DQD$_2$ into resonance using its charge detuning parameter $\delta_2$. We observe the transition of a single qubit vacuum Rabi mode splitting spectrum at large detunings $\delta_2$, to a well-resolved two-qubit vacuum Rabi mode splitting spectrum [Fig.~\ref{fig2:OnResonance}(d)] at $\delta_2=0$, with the collectively enhanced two-qubit coupling rate $g_\mathrm{c}/2\pi=\sqrt{g_1^2+g_2^2}/2\pi=$~77~MHz [Fig.~\ref{fig2:OnResonance}(c)]. This is a clear signature of the coherent photon-mediated coupling between two spatially separated DQDs in the resonant regime.

On resonance, the three systems (r3) form a triplet of two bright states $|\pm\rangle_{\rm{r3}}=(g_2|g,e,0\rangle+g_1|e,g,0\rangle\pm g_c|g,g,1\rangle)/\sqrt{2}g_c$ and one dark state $|0\rangle_{\rm{r3}}=(g_1|g,e,0\rangle-g_2|e,g,0\rangle)/g_c$ at frequencies $\omega_{|\pm\rangle_\text{r3}}=\omega_\mathrm{r}\pm g_c$ and $\omega_{|0\rangle_\text{r3}}=\omega_\mathrm{r}$, see schematic in Fig.~\ref{fig2:OnResonance}(g) and Appendix~\ref{sec:Model_sys}. This feature occurs because the drive field acts symmetrically on both qubits exciting only the symmetric qubit superposition of the bright states but not the anti-symmetric superposition of the dark state \cite{Majer2007,Fink2009}.
The data in Fig.~\ref{fig2:OnResonance}(d) shows excellent quantitative agreement with the master equation model see dashed red lines indicating the transition frequencies between the ground and the joint excited states allowing us to extract all relevant system parameters [Figs.~\ref{fig2:OnResonance}(d) and (e)].


Alternatively, coherent coupling between spatially separated DQDs can be mediated by virtual photons when transitions of two DQDs are resonant with each other but detuned from the resonator by $\Delta_\mathrm{r}=\omega_\mathrm{r}-\omega_\mathrm{DQD}$. In this case the effective coupling strength is reduced but the coupling mechanism is insensitive to photon loss from the resonator.



\begin{figure}[t!]
\includegraphics[width=0.50\textwidth]{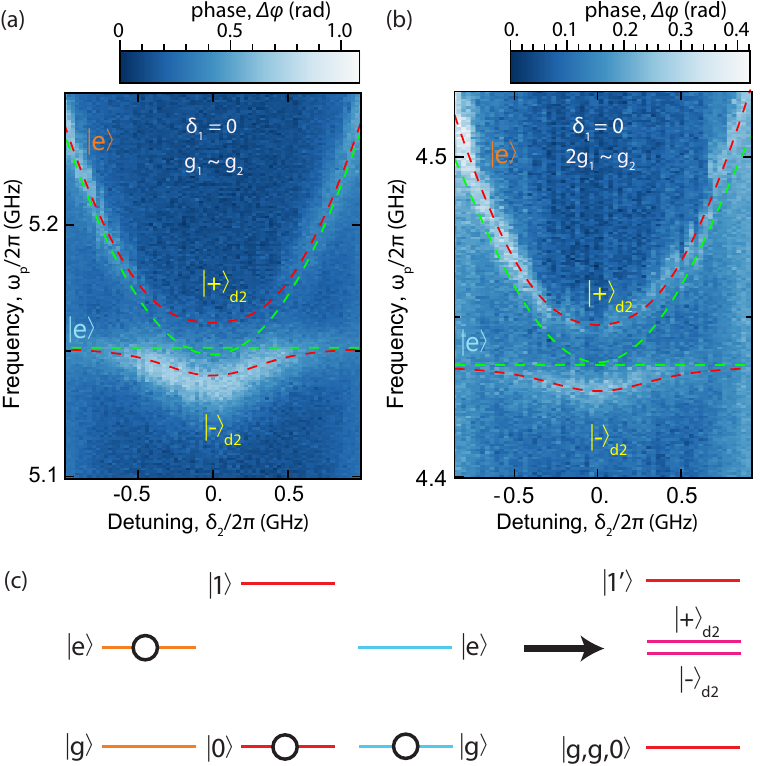}
\caption{Qubit spectroscopy of the virtual photon-mediated qubit-qubit exchange interaction.
(a) Resonator phase shift $\Delta\varphi$ for DQDs with equal coupling ($g_1\sim g_2$) at fixed $2t_1/2\pi=5.156$~GHz, $\delta_1=0$ and $2t_2/2\pi=5.148$~GHz \emph{vs.} detuning parameter $\delta_2$. Linetraces and additional data are in Appendix~\ref{sec:2J_equal}.
(b) Qubit spectroscopy for DQDs with different coupling ($2 g_1 \sim g_2$) and $\omega_\mathrm{DQD1}/2\pi(\delta_1=0)=4.436$~GHz measuring the phase shift $\Delta\varphi$ of the resonator \emph{vs.}~$\delta_2$ for, $2t_2/2\pi=4.443$~GHz.
(c) Schematic energy level diagram of two DQDs in resonance and detuned resonator. Basis states are shown on the left, hybridized states on the right with labels discussed in text and in Appendix~\ref{sec:Model_sys}.
}
\label{fig3:Dispersive_2J}
\end{figure}

To observe the dispersive coupling, we tune the resonator to $\omega_\text{r}/2\pi = 5.454$~GHz resulting in a detuning $\Delta_\mathrm{r}/2\pi\approx300$~MHz when both qubits are at $\delta_{k}=0$. The virtual photon-mediated exchange coupling is observed by the formation of a dark and bright state split in energy [Fig.~\ref{fig3:Dispersive_2J}(a)] when the two DQDs are (approximately) in resonance $\omega_\text{DQD1} \sim \omega_\text{DQD2}$. This also allows us to identify the resonances as transitions to the dispersively coupled two qubit entangled states  $|\pm\rangle_{\rm{d2}}=(g_1|g,e\rangle\mp g_2|e,g\rangle)/g_\mathrm{c}$~\cite{Majer2007} [Fig.~\ref{fig3:Dispersive_2J}]. Due to the near-equal coupling rates, $g_1\sim g_2$, the dark state is fully developed when the DQDs are resonant, $\omega_\text{DQD1}=\omega_\text{DQD2}$ (see Appendix~\ref{sec:Model_sys}). Then only the bright state is directly observable in qubit spectroscopy, see line traces in Appendix~\ref{sec:2J_equal}. To observe the splitting directly we can instead bias the DQDs to achieve $g_1\neq g_2$ which through the asymmetry in parameters makes the otherwise dark state observable.

We therefore configure both DQDs at a new charge bias point at which the coupling rates to the resonator are $g_1/2\pi=$~34~MHz and $g_2/2\pi=$~69~MHz (see Appendix~\ref{sec:res:uneq}). At $\delta_{1,2}\sim 0$, the qubit linewidths $\Gamma_{2,1}/2\pi=(4.6\pm0.6)$~MHz and $\Gamma_{2,2}/2\pi=(6.3\pm1.1)$~MHz are determined from spectroscopy measurements with the other qubit largely detuned. For these measurements, the resonator is tuned to $\omega_\text{r}/2\pi = 4.717$~GHz ($\kappa_\mathrm{int}/2\pi=8$~MHz, $\kappa_\mathrm{ext}/2\pi=4$~MHz).

Next we perform qubit spectroscopy with $2t_1/2\pi\sim2 t_2/2\pi\sim4.44$~GHz, corresponding to a detuning $\Delta_\mathrm{r}/2\pi\approx280$~MHz of each DQD from the resonator at $\delta_{1,2}=0$, putting the system in the dispersive regime~\cite{Wallraff2004}. A virtual photon-mediated exchange coupling $2J/2\pi =24.8$~MHz is observed spectroscopically when varying the detuning $\delta_2$ and keeping the bias parameters of DQD$_1$ fixed [Fig.~\ref{fig3:Dispersive_2J}(b)]. The system parameters which are used to display the non-interacting transition frequencies (green dashed lines) in Figs.~\ref{fig3:Dispersive_2J}(a) and (b) are extracted from a master equation simulation (red dashed lines).

We note that the spectroscopic lines of DQD$_1$ at large detuning $|\delta_2/2\pi|\gtrsim0.5$~GHz are less pronounced, due to its weaker coupling to the resonator and because DQD$_2$ is dispersively shifting the resonator, rendering read-out less sensitive for DQD$_1$.

\begin{figure}[t!]
\includegraphics[width=0.5\textwidth]{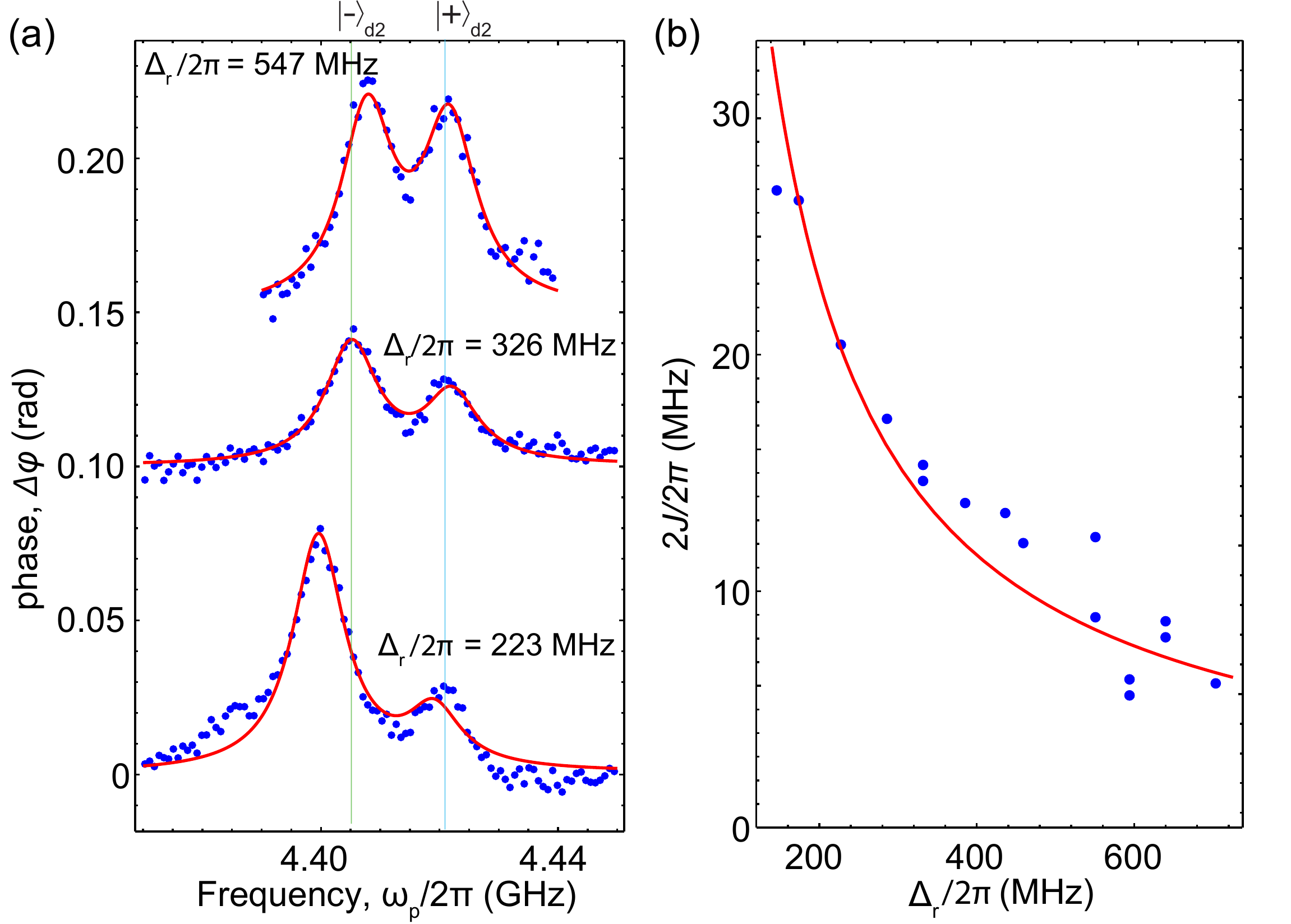}
\caption{Coherent qubit-qubit exchange splitting $2J$ \emph{vs.} resonator detuning $\Delta_\mathrm{r}(\Phi)$.
(a) Resonator phase shift (data points offset for clarity) measurement performed to extract qubit-qubit interaction $2 J$ on resonance for the indicated detunings $\Delta_\mathrm{r}(\Phi)$. Solid lines are fits to master equation simulations, see Appedices~\ref{sec:Model_sys} and \ref{sec:para} for details and parameters.
(b) $2J$ extracted from (b) and similar data 
\emph{vs.}~$\Delta_\text{r}$. Solid line is a fit to $1/\Delta_\text{r}$.}
\label{fig4:Dispersive_2J_detuning}
\end{figure}

Finally, we determine the scaling of the exchange coupling $J$ with detuning $\Delta_\text{r}$ from the flux-tuned resonator for the same fixed qubit parameters at $\delta_{1,2}=0$ [Fig.~\ref{fig4:Dispersive_2J_detuning}(b)]. We use the configuration of Fig.~\ref{fig3:Dispersive_2J}(b) since both symmetric and anti-symmetric resonances are observable for $\omega_\text{DQD1}=\omega_\text{DQD2}$ which allow to extract $2J$ by fitting to results of the the master-equation simulation [Fig.~\ref{fig4:Dispersive_2J_detuning}(a)]. At small resonator detunings we find the largest coherent qubit-qubit exchange rates of $2J/2\pi=27$~MHz [Fig.~\ref{fig4:Dispersive_2J_detuning}(b)] clearly exceeding the combined qubit linewidths $(\Gamma_{2,1}+\Gamma_{2,2})/2\pi=11$~MHz. For $\Delta_\text{r}/2\pi>$~560~MHz~$ \sim 8g_2/2\pi$ the $2J$ is smaller than the qubit linewidths.

We note that the transition of the dark state $|+\rangle_\text{d2}$ remains at fixed frequency while the bright state $|-\rangle_\text{d2}$ shifts as function of $\Delta_\text{r}$ in agreement with our master equation model [Fig.~\ref{fig4:Dispersive_2J_detuning}(a)], see Appendix~\ref{sec:Model_sys}. When plotting $2J$ \emph{vs.} the resonator detuning, $\Delta_\mathrm{r}\left(\Phi\right)$, we find approximately the expected scaling with $1/\Delta_\mathrm{r}$ [Fig.~\ref{fig4:Dispersive_2J_detuning}(b)]. Also the overall coupling strength $g_1g_2/4\pi^2=$~2.1$\cdot10^3$~MHz$^2$ is consistent with the one calculated 2.4$\cdot10^3$~MHz$^2$ from the individually measured qubit-resonator coupling rates, $g_{1,2}$ and the detuning, $\Delta_\text{r}$. 




We emphasize that the device investigated here features a frequency tunable resonator ($\omega_\text{r}$), charge qubits with tunable transition frequency ($\delta$), sweet-spot ($2t$) and dipole coupling strength ($g$) enabling a comprehensive study of coherent photon-mediated coupling phenomena, the concepts of which are transferable to other semiconductor material systems. We also point out that  photon-mediated coupling enables two-qubit gates between charge or spin qubits across micrometer, millimeter or even longer distances which is essential for scaling quantum information processing with semiconductor qubits \cite{Childress2004,Vandersypen2017}. In superconducting circuits the observation of long-range qubit-qubit coupling \cite{Majer2007,Fink2009} led to the development of both resonant and dispersive photon-mediated two-qubit gates \cite{Sillanpaa2007,DiCarlo2009} and enabled the scaling of circuits to the level of several tens of qubits \cite{Otterbach2017}.





\appendix

\addtocontents{toc}{\setcounter{tocdepth}{0}}

\section{Device and fabrication}\label{sec:fab}

The substrate is a commercially available GaAs wafer with a 500~nm GaAs layer grown by molecular beam epitaxy. Subsequently, a 40~nm layer of Al$_\mathrm{x}$Ga$_\mathrm{1-x}$As is grown as a spacer to a $\delta$-donor layer of silicon dopants followed by 45~nm of Al$_\mathrm{x}$Ga$_\mathrm{1-x}$As and capped by 5~nm GaAs layer. 90~nm below the surface, at the interface of GaAs/Al$_\mathrm{x}$Ga$_\mathrm{1-x}$As, a 2-dimensional electron gas (2DEG) is formed by bending the conduction band below the Fermi-energy. 

\begin{figure}[t!]
\begin{center}
\includegraphics[width=0.5\textwidth]{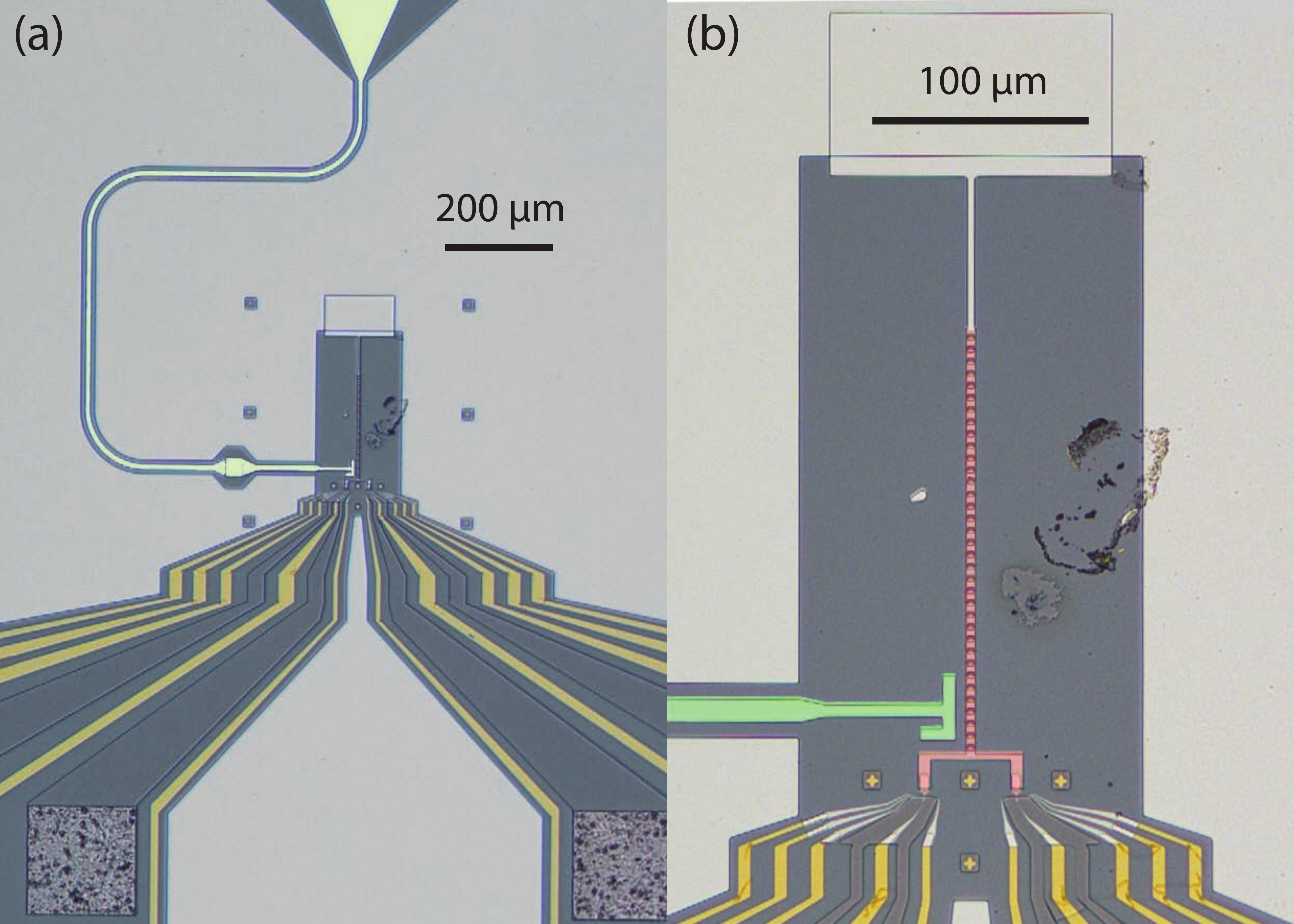}
\end{center}
\caption{Overview images of the device with only the drive line (green) and SQUID array resonator (red) false coloured. (a) Overview optical image of the device, showing at the top the launcher of the drive line and its waveguide (green). At the bottom, the gate lines (yellow) are visible. The two squares at the bottom left and right are the ohmic contacts to the 2DEG. (b) Enlarged view of (a)}
\label{fig:images}
\end{figure}

In the first photolithography step a GaAs mesa hosting the DQDs is formed by wet-etching with a Piranha solution. The source and drain up to the DQDs are also formed as part of the mesa. We emphasize that all 2DEG is removed below the SQUID array resonator to maintain its quality factor. The fabrication residue visible in the resonator area, in Fig.~\ref{fig:images}, could not be removed. We believe that this residue did not lead to any reduction in device performance as the internal loss rate of the resonator is similar in previous devices~\cite{Stockklauser2017,Scarlino2017b}. In the subsequent photolithography lift-off step, the ohmic contact of the DQD source and drain are deposited by electron beam evaporation of a Ge/Au/Ni layer, which are annealed at 470$^\circ$C for 5~minutes to diffuse into the 2DEG layer.

The electrostatic gates are created in two lithography steps. First the coarse gates and pads [yellow/gold structures in Figs.~\ref{fig1:SampleAndCircuit}(b) and (c)] are patterned with optical lithography and Ti/Au (5/80~nm) is deposited by electron beam evaporation and then lifted-off. At this step the markers for the electron beam pattering, the gold (yellow) crosses, visible in Figs.~\ref{fig1:SampleAndCircuit}(c) and~\ref{fig:images}, are also deposited. The finer structures of the gates are done in a subsequent step.

The ground plane is defined in the last step of optical lithography. The drive line is patterned in this step up to a distance of 200~$\mu$m from the resonator, see Fig.~\ref{fig:images}(a). The ground plane is made of Ti/Al (3/200~nm) by lift-off and is deposited by electron beam evaporation. A part of the ground plane, the light grey areas, are visible in Fig.~\ref{fig:images}.

The first electron beam lithography step defines the fine gates in a PMMA mask for lift-off, using 3/25nm (Ti/Al) deposited by electron beam evaporation. The resulting fine gates are shown in Fig.~\ref{fig1:SampleAndCircuit}(e).

In the final step, a PMMA/MMA bilayer resist is patterned with electron beam lithography. The Dolan-bridge technique~\cite{Dolan1977} - the angle evaporation of two Al layers (35/110~nm) interrupted by an oxidation step - is used to create the Josephson junctions for the SQUID array resonator which is connected to the ground plane and plunger gate of both DQDs. In addition, the resonator drive line is deposited in the same step to assure good alignment between the drive line and the resonator defining the coupling capacitance and thus the coupling rate, $\kappa_\mathrm{ext}$. The drive line splits the ground plane which is reconnected by multiple wirebonds, (not show in Fig.~\ref{fig:images}).

The device is bonded in a PCB and mounted in a Oxford Triton 200 cryofree dilution refrigerator at the base plate with a typical temperature of $\sim 20$~mK~\cite{Stockklauser2017b}.

\section{Measurement setup}\label{sec:setup}

The SQUID array resonator is measured in reflection by applying a microwave tone at the drive line (green in Fig.~\ref{fig:images}). The microwave tone is generated at room-temperature and is attenuated (-20~dB) at the 4~K, 100~mK and 20~mK stages before passing through a circulator which routes it to the resonator and routes the reflected signal to the output line. In the output line the reflected signal is amplified using a Low Noise Factory HEMT (+39~dB) at 4~K and two amplifiers (+33~dB each) at room-temperature, before it is down converted to an intermediated frequency (IF) of 250~MHz. With +29~dB amplification the IF signal is acquired at 1Gs/s using an Acqiris U1084A PCIe 8-bit High-Speed Digitizer.

The DC voltages to the gates are supplied by Yokogawa 7651 DC programmable sources with a 1:11 voltage divider also acting as a low pass filter (1~Hz cut-off). The source and drain of both DQDs where grounded in the experiment. At base temperature, 2-stage RC filters with 16~kHz and 160~kHz cut-off are used at the input of shielded lines leading to the sample holder.

\begin{figure}[ht]
\begin{center}
\includegraphics[width=0.5\textwidth]{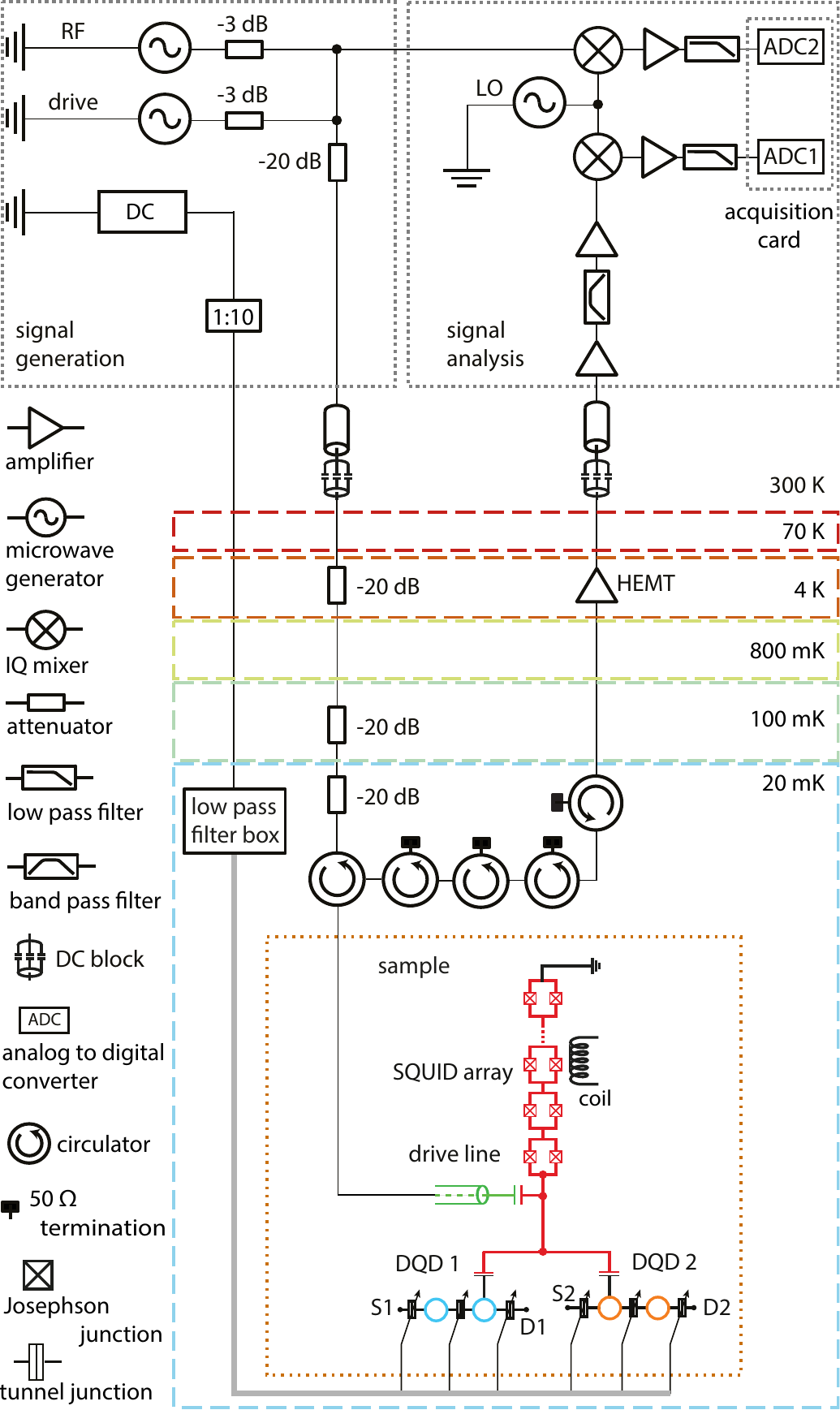}
\end{center}
\caption{Simplified schematic of the cryogenic and room-temperature components and equipment used in the experiments, further details are provided in text.}
\label{fig:Simplified_circuit}
\end{figure}

A schematic of the complete setup with all important components is displayed in Fig.~\ref{fig:Simplified_circuit}.

\section{Modelling the system}\label{sec:Model_sys}

Here we discuss the theoretical model used for understanding and fitting the experimental data.
We describe the system using the Hamiltonian
\begin{align}
	\H_{\text{tot}} = \H_\text{\text{res}} + \sum_{k} \H_{k} + \sum_{k} \H_{\text{int},k} \,,
\end{align}
with the resonator Hamiltonian
\begin{align}
	\H_{\text{res}} =  \omega_{\text{r}} a\hc a \,,
\end{align}
the Hamiltonian for the $k$-th DQD ($k=1,2$ for the experiments discussed here)
\begin{align}
	\H_{k} = -\frac12  \delta_{k} \sigma_{z} + t_{k} \sigma_{x} \,,
\end{align}
and the coupling between resonator and DQDs,
\begin{align}
	\H_{\text{int},k} =  g_{k} \sigma_{z} (a\hc + a)\,.
\end{align}
Here we use $\hbar=1$ for simplicity. $a$ ($a^\dagger$) is the bosonic annihilation (creation) operator, $\omega_{\text{r}}$ is the resonator angular frequency, 
and $\sigma_{x,y,z}$ are Pauli matrices.
The DQDs are defined through the charge detuning $\delta_k$ and their tunnel splitting $2t_k$. The parameter $g_{k}$ is the dipolar coupling strength between the $k$-th DQD and the resonator.
Transforming $\H_{\text{tot}}$ into the DQD eigenbasis, we find
\begin{align}
	\H =&  \omega_\text{r} a\hc a -\frac12 \sum_{k}  \omega_{k} {\sigma}_{z} \nn\\
			 &+  g_{k} \sum_{k} \l( \sin{\theta_{k}} {\sigma}_{x} + \cos{\theta_{k}} {\sigma}_{z} \r) (a\hc + a) \,,\nn\\
		\approx&  \omega_\text{r} a\hc a -\frac12 \sum_{k} \omega_{k} {\sigma}_{z} + g_{k} \sum_{k} \sin{\theta_{k}} \l({\sigma}_{-} a\hc + {\sigma}_{+} a \r) \,,
\label{eq:HTot}
\end{align}
where we performed a rotating wave approximation in the last step to arrive at the well-known Tavis-Cummings Hamiltonian.
Here $\omega_{k} = \sqrt{4t_{k}^{2} + \delta_{k}^{2}}$
and $\tan{\theta_{k}} = 2t_{k} / \delta_{k}$, so that $\sin{\theta_{k}} = 2t_{k} / \omega_{k}$ and $\cos{\theta_{k}} = \delta_{k} / \omega_{k}$.

\subsection{Scattering in input-output theory}

To model reflection of an incident signal from the resonator, we use the Scattering Lindblad Hamiltonian (SLH) cascaded quantum systems formalism~\cite{Combes2016a}.
Modelling the cavity as a single port resonator with output coupling $\kappa_\text{ext}$ and cascading in the coherent probe at signal frequency $\omega_\text{p}$~\cite{Muller:PRA:2017}, we find the total SLH Hamiltonian
$\H_{\text{SLH}} = \H_{\text{tot}} + \H_{\text P} $, where the probe term is
\begin{align}
	\H_{\text P} = \frac{1}{2\ii} \sqrt{\kappa_{\text{ext}}} \l( \alpha a\hc - \alpha\cc a  \r) \,,
\end{align}
with $\alpha$, the input coherent field amplitude and $\kappa_{\text{ext}}$, the radiative coupling to the external waveguide modes.
The probe Hamiltonian $\H_{\text{P}}$ is already written in a frame rotating at the signal frequency $\omega_\text{p}$.
We transform $\H_{\text{tot}}$ into the same rotating frame and find
\begin{align}
	\H_{\text{tot}} =&
		 \delta\omega_\text{r} a\hc a -\frac12 \sum_{k}  \delta\omega_{k} {\sigma}_{z}^{(k)} \nn\\
		 	&+ g_{k} \sum_{k} \sin{\theta_{k}} \l({\sigma}_{-}^{(k)} a\hc + {\sigma}_{+}^{(k)} a \r)\,,
\end{align}
with the detunings $\delta\omega_\text{r} = \omega_\text{r} - \omega_\text{p}$ and $\delta\omega_{k} = \omega_{k} - \omega_\text{p}$. Including incoherent processes, the time evolution of the systems density matrix $\rho$ follows the master equation
\begin{align}
	\dot \rho = -\ii \comm{\H_{\text{SLH}}}{\rho} + \mathcal L_{\text{nr}}\rho + \mathcal L_{\text{SLH}}\rho\,.
	\label{eq:MESLH}
\end{align}
The second term on the RHS of the master equation~\eqref{eq:MESLH} describes all non-radiative losses and dephasing processes.
For zero-temperature quantum baths coupled to each quantum dot and the resonator independently, we write this as
\begin{align}
	\mathcal L_{\text{nr}}\rho = \sum_{k} \gamma_{1,k} \diss{{\sigma}_{-}}\rho + \frac12 \sum_{k} \gamma_{\varphi,k} \diss{{\sigma}_{z}} \rho + \kappa_{\text{in}} \diss{a}\rho \,, \label{eq:dephase}
\end{align}
with the DQDs relaxation rate $\gamma_{1,k}$, their pure dephasing rates $\gamma_{\varphi,k}$ and the internal resonator decay into non-guided modes $\kappa_{\text{in}}$.

Here we assume 
that the main loss channels for the DQDs is a coupling to electromagnetic modes of the environment described by the dipole operator
\begin{align}
	H_{\text{env}} = {\sigma}_{z} \sum_\text{k} \beta_\text{k} (b_\text{k} + b_\text{k}\hc) \,,
\end{align}
where $b_\text{k}$ ($b_\text{k}^\dagger$) are bosonic annihilation (creation) operators for a mode of the electromagnetic environment to DQD-k. We can find the DQD relaxation and dephasing rates~\cite{Muller:PRA:2016}
\begin{align}
	\gamma_\text{1,k} &= \sin^{2}{\theta_\text{k}} C(\omega_\text{k}) \,,\nn\\
	\gamma_{\varphi\text{,k}} &= \cos^{2}{\theta_\text{k}} C(0) \,,
\end{align}
where $C(\omega_k)$ is the environmental spectral function, $C(\omega) = \int dt e^{-i \omega t} \left< \hat X(t) \hat X(0)\right>$, with $\hat X= \sum_\text{k} \beta_\text{k} (b_\text{k} + b\hc_\text{k})$.
In our calculations we assumed white noise spectra for the noise acting on the DQDs for simplicity.

Finally the third term  on the RHS of Eq.~\eqref{eq:MESLH} describes the scattering of the input drive fields into the waveguide modes as
\begin{align}
	\mathcal L_{\text{SLH}}\rho =  \diss{L} \rho \,,
\end{align}
where
\begin{align}
	L &= \sqrt{\kappa_{\text{ext}}} \: a + \alpha \mathds{1}.
\end{align}
We calculate the amplitudes $\beta$ and photon fluxes $n$ of the scattered fields from
\begin{align}
	\beta = \Tr{\l\{L\rho\r\}} \quad , \quad n = \Tr{\l\{L\hc L \rho\r\}}
\end{align}
where $\rho$ is the solution of the master equation Eq.~\eqref{eq:MESLH}.
For spectroscopy experiments, as modelled here, it is sufficient to calculate the steady-state scattering, considering $\dot \rho = 0$.

\subsection{Two-tone spectroscopy}\label{sec:TwoTone}

In principle, the technique described here would allow us to simulate circuit QED spectroscopy~\cite{Blais2004} directly, by either adding another set of input and output modes at different frequencies or,
assuming that the input at or close to the DQD resonance is not monitored, by adding a coherent drive term to the Hamiltonian.
Since there is now multiple time-dependent terms in the Hamiltonian, which oscillate at different frequencies, a single rotating frame is no longer sufficient to capture the dynamics.
Instead one can move towards a multi-tone Floquet analysis or alternatively perform time-dependent simulations of the dynamics to find the response of the system.

In practice this has proven not feasible as the number of unknown parameters is too large for reliable fits to the data.
We have therefore fitted the qubit spectroscopy experiments with simulations of standard single-tone spectroscopy in the far detuned regime, adding an additional scale and offset parameter to match the amplitude of the experimental results. We stress that the relative height of the resonances is extracted directly from the master equation, see e.g. Figs.~\ref{fig:SI:symmetric_g_2J}(d) and (h).
Since the two-tone experiments are in the linear response regime of the resonator phase (weak drive and read-out power), i.e. the change in the signal phase is linear in the excitation probability of the DQDs,
this technique can still produce quantitative agreement with the experimental data.

%

\subsection{Eigenstates in the coupled system}

To clarify the composition of the eigenstates at the points of maximal coupling, we present here the exact expressions for the two cases where
\begin{itemize}
	\item[(a)]{both DQDs and the resonator are resonant, $\Delta_\mathrm{r}=\omega_\text{r}-\omega_{k}=0$, relevant for Fig.~\ref{fig2:OnResonance} of the main text, and}
	\item[(b)]{the two DQDs are resonant with each other and the resonator is detuned, $\Delta_\mathrm{r}\gg g_k$, relevant for Fig.~\ref{fig3:Dispersive_2J} and \ref{fig4:Dispersive_2J_detuning} of the main text.}
\end{itemize}

\begin{figure*}
\begin{center}
\includegraphics[width=1.0\textwidth]{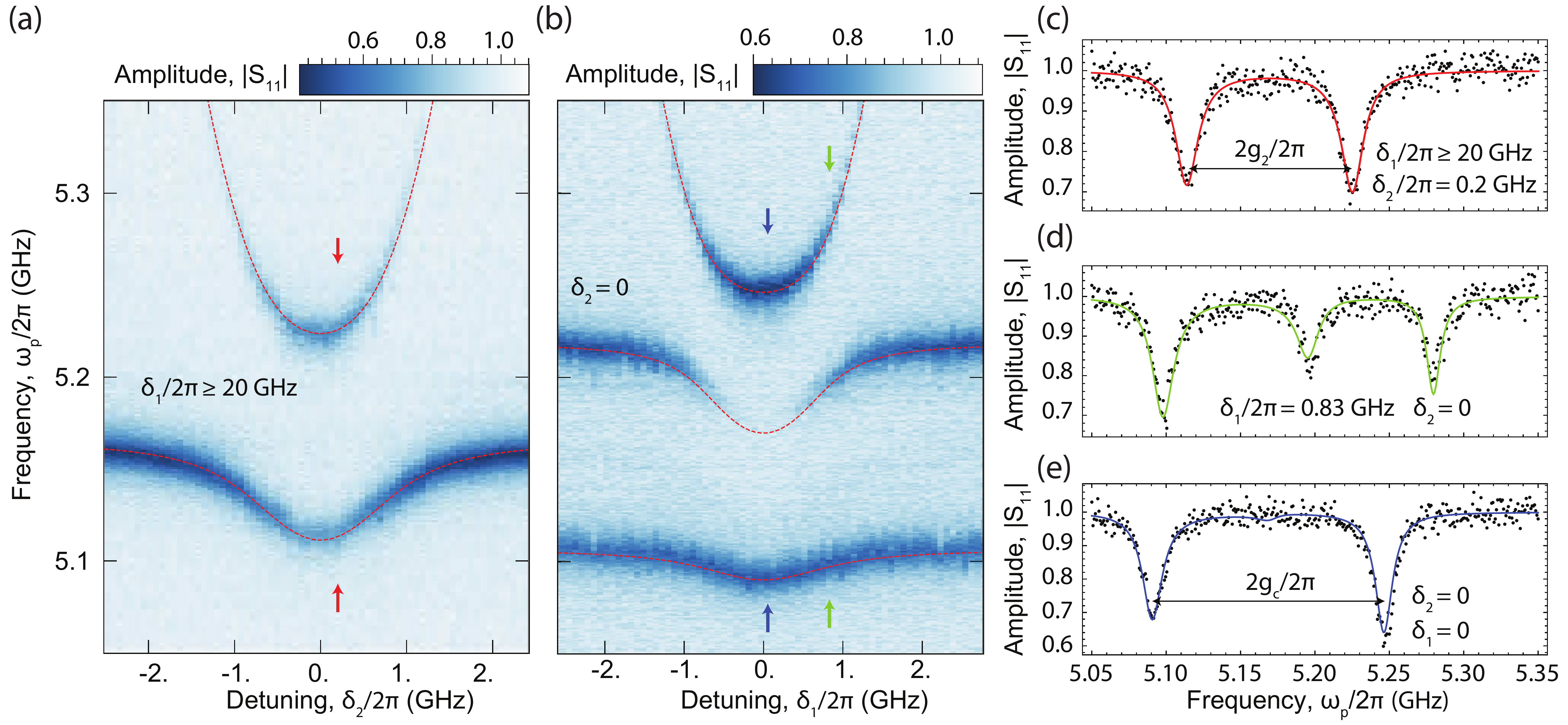}
\end{center}
\caption{Resonator spectroscopy with one or two DQDs tuned into resonance with the resonator. Same charge configuration as used for the data in Fig.~\ref{fig2:OnResonance} of the main text. (a) Tuning $\delta_2$ of DQD$_2$ ($2t_2/2\pi=5.168$~GHz) into resonance with the SQUID array resonator, when DQD$_1$ is tuned in Coulomb blockade, $\delta_\mathrm{1}/2\pi\gtrsim20$~GHz. (b) With, $\omega_\mathrm{r}=\omega_\mathrm{DQD2}$, DQD$_1$ ($2t_1/2\pi=5.183$~GHz) is brought in resonance by changing, $\delta_1$. Panels (c-e) are linetraces of panels (a) and (b) to show the conditions, $\omega_\mathrm{r}=\omega_\mathrm{DQD2}$ (red arrows), $\omega_\mathrm{DQD2}+g_1\approx\omega_\mathrm{DQD1}$ (green arrows) and $\omega_\mathrm{r}=\omega_\mathrm{DQD2}=\omega_\mathrm{DQD1}$ (blue arrows). All fits in panel (c-e) are fits to master equation with extracted parameters listed and discussed in Appendix~\ref{sec:para}.}
\label{fig:SI:symmetric_g_resonance}
\end{figure*}

\begin{figure*}
\begin{center}
\includegraphics[width=1.0\textwidth]{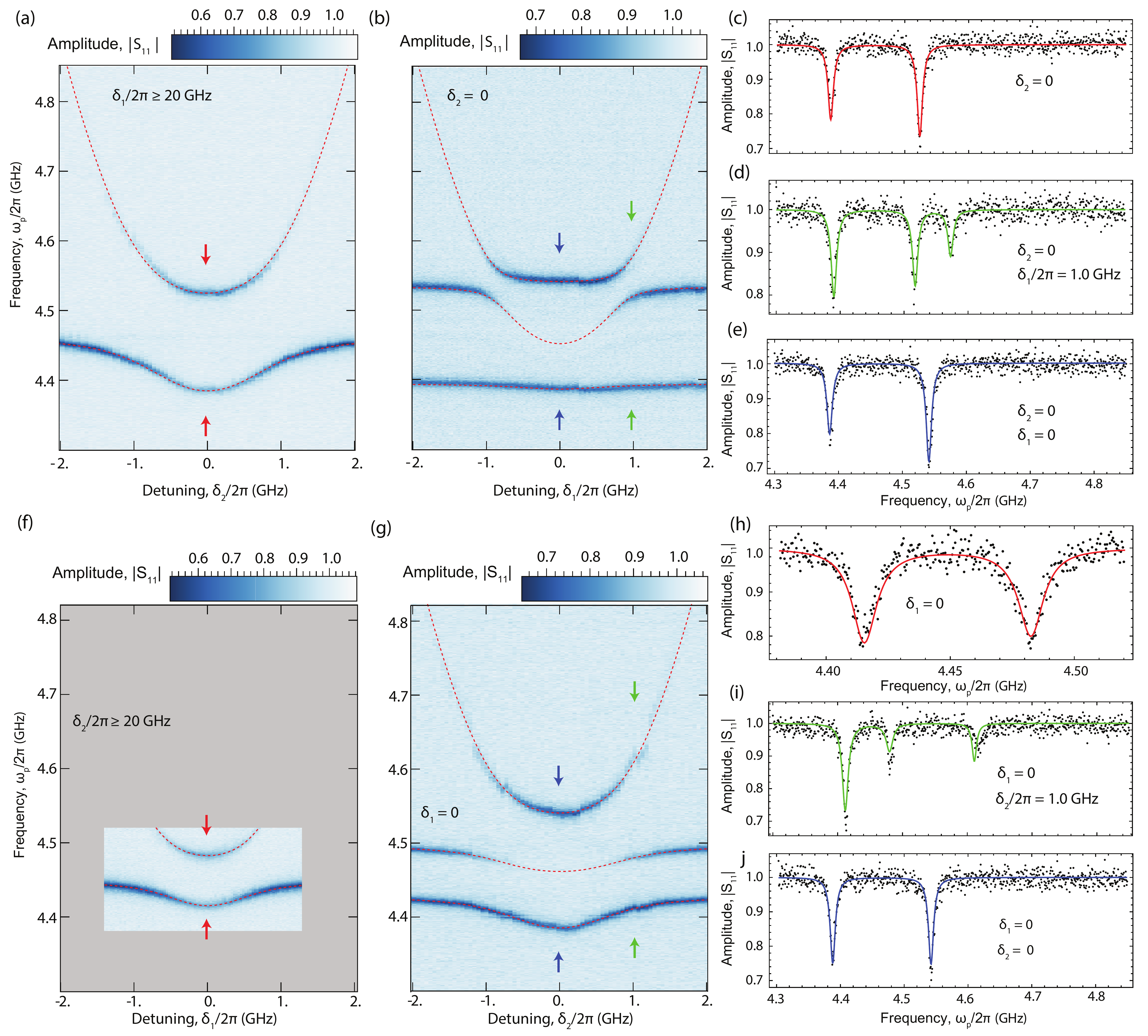}
\end{center}
\caption{Resonator spectroscopy with one or two qubits tuned into resonance with the resonator. Similar to data in Fig.~\ref{fig2:OnResonance} except that the device is biased at a point with coupling rates $g_1/2\pi=34$~MHz and $g_2/2\pi=69$~MHz. All data is fitted with master equation simulations (Appendices~\ref{sec:Model_sys} and~\ref{sec:para}). (a) Measuring the reflection spectrum $|S_{11}|$, while tuning $\delta_2$ of DQD$_2$ ($2t_2/2\pi=4.448$~GHz) and DQD$_1$ is in Coulomb blockade $\delta_1/2\pi\gtrsim20$~GHz. The resonator is flux tuned to $\omega_\mathrm{r}/2\pi=4.462$~GHz. (b) Reflection spectrum $|S_{11}|$, when resonator - DQD$_2$ are in resonance ($\delta_2=0$), the detuning $\delta_1$ of the weaker coupled DQD$_1$ ($2t_1/2\pi=4.452$~GHz) is swept. (c) Linetrace of reflection spectrum $|S_{11}|$ in (a) at the red arrows showing two separate resonances. (d) Resonator spectroscopy line trace obtained from (b) when $\omega_\mathrm{DQD2}+g_2\approx\omega_\mathrm{DQD1}$ (green arrows). (e) Resonator spectroscopy line trace adopted from panel (b) at $\delta_{1,2}=0$ obtaining $\omega_\mathrm{r}=\omega_\mathrm{DQD1}=\omega_\mathrm{DQD2}$ (blue arrows). (f) Reflection spectrum $|S_{11}|$, while tuning $\delta_1$ of DQD$_1$ ($2t_1/2\pi=4.450$~GHz), when DQD$_2$ is in Coulomb blockade $\delta_2/2\pi\gtrsim20$~GHz. (g) Keeping DQD$_1$ - resonator in resonance, DQD$_2$ ($2t_2/2\pi=4.461$~GHz) is tuned \emph{vs.} $\delta_2$, resulting in hybridization in the case where all three systems are in resonance, $\omega_\mathrm{r}=\omega_\mathrm{DQD2}=\omega_\mathrm{DQD1}$. (h) Line trace of panel (f) (red arrows) showing the reflection spectrum $|S_{11}|$ with two resonances. (i) Line trace of panel (g) (green arrows) taken at $\omega_\mathrm{DQD1}+g_1\approx\omega_\mathrm{DQD2}$, showing the DQD$_2$ tuned into resonance with the
hybridized DQD$_1$~-~resonator state, $|+\rangle_{r3}$. (j) Data taken at the similar configuration as, (e), $\delta_{1,2}=0$, linetrace of the $|S_{11}|$ in panel (g) (blue arrows).}
\label{fig:SI:opposite_fig2_main}
\end{figure*}

In case (a) the Hamiltonian in the one-excitation subspace can be written as
\begin{align}
	H_{(a)} = \l( \begin{array}{ccc}
			0 & g_{1} & 0 \\
			g_{1} & 0 & g_{2}  \\
			0 & g_{2} & 0
		\end{array} \r) \,,
\end{align}
where we subtracted a constant energy offset, $\omega_k =\omega_\text{r}$, and we are considering the basis $\l\{ \ket{e,g,0}, \ket{g,g,1}, \ket{g,e,0} \r\}$.
Diagonalising this Hamiltonian leads to the eigenstates and eigenenergies
\begin{eqnarray}
	E_{0} = 0 \,, \quad 	E_{-} = -g_\mathrm{c} \,, \quad E_{+} = + g_\mathrm{c} \,, \nn
\end{eqnarray}
and corresponding eigenstates
\begin{align}
	\ket{0}_\text{r3} &= \frac{1}{g_\mathrm{c}} \l( g_{1} \ket{g,e,0} -g_{2} \ket{e,g,0} \r) \,,\nn\\
	\ket{-}_\text{r3} &= \frac{1}{\sqrt{2}g_\mathrm{c}} \l( g_{2} \ket{g,e,0} + g_{1} \ket{e,g,0} - g_\mathrm{c} \ket{g,g,1} \r) \,, \nn\\
	\ket{+}_\text{r3} &= \frac{1}{\sqrt{2}g_\mathrm{c}} \l( g_{2} \ket{g,e,0} + g_{1} \ket{e,g,0} + g_\mathrm{c} \ket{g,g,1} \r) \,, \nn
\end{align}
with $g_\mathrm{c} = \sqrt{g_{1}^{2} + g_{2}^{2}}$. Here the state $\ket{0}_\text{r3}$ is a dark state with respect to the coupling to the resonator as it is an anti-symmetric state and the coupling between DQDs and resonator is symmetric, since both DQDs couple to the same phase of the drive field at one end of the resonator.
The condition (a) is exactly met when, $\omega_\mathrm{r}=\omega_\mathrm{DQD1}=\omega_\mathrm{DQD2}$, as illustrated by the data in Figs.~\ref{fig2:OnResonance}(b) and (c), \ref{fig:SI:symmetric_g_resonance}(b), (e), (g), (j) and \ref{fig:SI:opposite_fig2_main}(b) and (e).

The second case (b) we treat here in two equivalent ways. First, we write the Hamiltonian
\begin{align}
	H_{(b1)} =  \l( \begin{array}{ccc}
			0 & g_{1} & 0 \\
			g_{1} & \Delta_\mathrm{r} & g_{2}  \\
			0 & g_{2} & 0
		\end{array} \r) \,,
	\label{eq:Hb}
\end{align}
where the only difference to $H_{(a)}$ is the non-zero energy of the resonator state compared to the DQD states, with $\Delta_\mathrm{r} = \omega_\text{r} - \omega_{k}\neq0$ and $\omega_k=0$.
Directly diagonalising this Hamiltonian is possible but the expressions for the eigenstates do not lend themselves to quick insights.
Instead we assume the relevant limit $\Delta_\mathrm{r}\gg g_{1}, g_{2}$, so that we can approximate
$\sqrt{\Delta_\mathrm{r}^{2} + 4 g_\mathrm{c}^{2}} \approx \Delta_\mathrm{r} \l( 1 + \frac{2g_\mathrm{c}^{2}}{\Delta_\mathrm{r}^{2}}  \r)$ and find in this limit
\begin{eqnarray}
	E'_{+} = 0 \,, \quad E'_{-} = -\frac{g_\mathrm{c}^{2}}{\Delta_\mathrm{r}}\,, \quad E'_{1} = \Delta_\mathrm{r} + \frac{g_\mathrm{c}^{2}}{\Delta_\mathrm{r}}\, ,
\end{eqnarray}
with the corresponding (unnormalized) eigenstates
\begin{eqnarray}
	\ket{+'}_\text{r3} &=& \frac{1}{g_\mathrm{c}} \l( g_{1} \ket{g,e,0} -g_{2} \ket{e,g,0} \r) \,,\nn\\
	\ket{-'}_\text{r3}  &=& \frac{1}{g_\mathrm{c}\sqrt{g_\mathrm{c}^{2}+\Delta_\mathrm{r}^{2}}} \times \nn\\
			&&\l( g_{2} \Delta_\mathrm{r} \ket{g,e,0} +   g_{1} \Delta_\mathrm{r} \ket{e,g,0} - g_\mathrm{c}^{2} \ket{g,g,1} \r)\nn\\
		&\approx& \frac{1}{g_\mathrm{c}} \l( g_{2} \ket{g,e,0} + g_{1} \ket{e,g,0} - \frac{g_\mathrm{c}^{2}}{\Delta_\mathrm{r}} \ket{g,g,1} \r) \,,\nn\\
	\ket{1'}_\text{r3} &=& \frac{1}{\sqrt{g_\mathrm{c}^{2} + \Delta_\mathrm{r}^{2}}} \l( g_{2} \ket{g,e,0} + g_{1} \ket{e,g,0} + \Delta_\mathrm{r} \ket{g,g,1} \r)\nn\\
		&\approx& \frac{1}{\Delta_\mathrm{r}} \l( g_{2} \ket{g,e,0} + g_{1} \ket{e,g,0} + \Delta_\mathrm{r} \ket{g,g,1} \r) \,,
	\label{eq:qubit_res_like_states}
\end{eqnarray}
with $g_\mathrm{c} = \sqrt{g_{1}^{2}+ g_{2}^{2}}$.
The states $\ket{\pm'}_\text{r3}$ are the qubit-like states and $\ket{1'}_\text{r3}$ is the resonator-like state used in the energy diagram in Fig.~\ref{fig3:Dispersive_2J}(c).
The difference in visibility of the $\ket{\pm}_{d2}$-states in Fig.~\ref{fig4:Dispersive_2J_detuning}(a) when changing the resonator detuning $\Delta_\mathrm{r}$, is full captured in this approximation.
The last term in $\ket{-'}_\text{r3}$ contains the excited state of the resonator and its coefficient is proportional to $1/\Delta_\mathrm{r}$ so that if we increase the resonator detuning
this coefficient and the visibility in spectroscopy decreases.
As is observed in Fig.~\ref{fig4:Dispersive_2J_detuning}(a), the visibility of the darker state $\ket +_\text{r3}$ remains constant while the brighter state $\ket -_\text{r3}$ becomes weaker.

In the same spirit, we may take the coupling $g_{k}$ as a perturbation, and, starting from the Hamiltonian Eq.~\eqref{eq:Hb}, find the approximate Hamiltonian for the DQDs in perturbation theory up to second order in $g_{k}/\Delta_\mathrm{r}$ as
\begin{align}
	H_{(b2)} = \frac{1}{\Delta_\mathrm{r}} \l( \begin{array}{cc}
			g_{1}^{2} & g_{1} g_{2} \\
			g_{1}g_{2} & g_{2}^{2}
		\end{array} \r) \,,
	\label{eq:H2}
\end{align}
in the basis $\l\{ \ket{e,g}, \ket{g,e} \r\}$. Note that here the two states are not in resonance, due to each states second-order energy correction obtained from the resonator dispersive shift.
This assumes that the DQDs are tuned such that, in absence of the resonator they would be resonant with each other.
The eigenvalues and eigenstates of this perturbative Hamiltonian are:
\begin{align}
	\bar E_{+} = 0 \quad , \quad \ket{+}_\text{d2} = \frac{1}{g_\mathrm{c}} \l(g_{1} \ket{g,e} - g_{2} \ket{e,g}\r) \,,\label{eq:dark}\\
	\bar E_{-} = -\frac{g_\mathrm{c}^{2}}{\Delta_\mathrm{r}} \quad , \quad \ket{-}_\text{d2} = \frac{1}{g_\mathrm{c}} \l( g_{2} \ket{g,e} + g_{1} \ket{e,g} \r)\,,
\end{align}
which is identical to Eq.~\eqref{eq:qubit_res_like_states} in the limit $\Delta_\mathrm{r}\gg g_{k}$.

\begin{figure*}
\begin{center}
\includegraphics[width=1.0\textwidth]{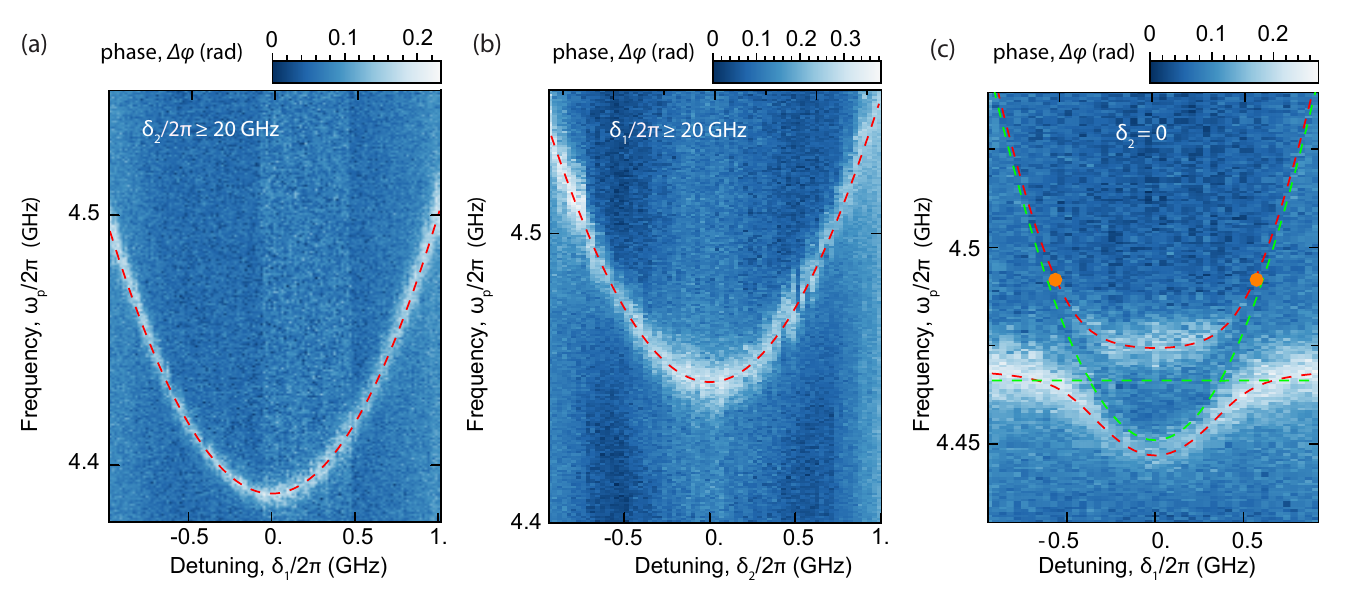}
\end{center}
\caption{Supplementary data to Fig.~\ref{fig3:Dispersive_2J}(b) showing qubit spectroscopy of the individual DQDs with the resonator tuned to $\omega_\mathrm{r}/2\pi=4.726$~GHz. Dashed lines are fits to extract relevant parameters. (a) Qubit spectroscopy of DQD$_1$ ($2t_1/2\pi=4.392$~GHz) \emph{vs.} detuning $\delta_1$ in the dispersive regime. (b) Qubit spectroscopy of DQD$_2$ ($2t_2/2\pi=4.448$~GHz) \emph{vs.} detuning $\delta_2$ in the dispersive regime. (c) Qubit spectroscopy measuring the resonator phase shift with DQD$_2$ fixed ($2t_2/2\pi=4.472$~GHz, $\delta_2=0$) and DQD$_1$ ($2t_1/2\pi=4.453$~GHz) tuned with $\delta_1$.}
\label{fig:SI:asymmetric_qubit_spec}
\end{figure*}

\begin{figure*}
\begin{center}
\includegraphics[width=1.0\textwidth]{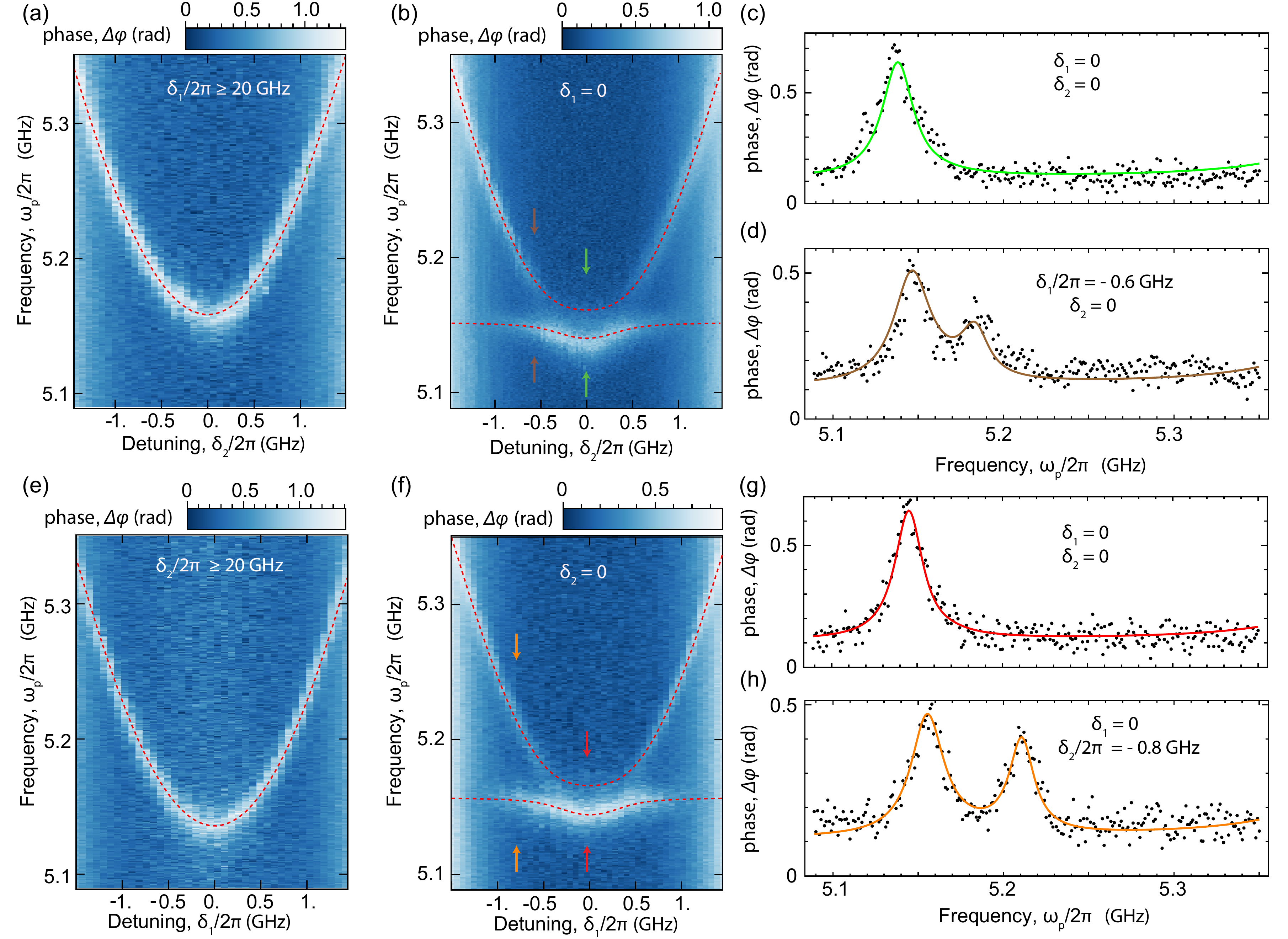}
\end{center}
\caption{Single and two qubit spectroscopy in the dispersive regime. The read-out is performed at 5.454~GHz by populating the resonator with $\sim$~0.3 photons on average, for all data shown. Red-dashed lines are fits to the master equation simulations and solid lines are master equation simulations, see Appendix~\ref{sec:para}. (a) Qubit spectroscopy of DQD$_2$ ($2t_2/2\pi=5.168$~GHz) \emph{vs.} detuning $\delta_1$. DQD$_1$ is detuned to, $\delta_1/2\pi\gtrsim 20$~GHz. (b) Qubit spectroscopy with DQD$_1$ at $\delta_1=0$ and performing qubit spectroscopy \emph{vs.} $\delta_2$. (c) Line trace at $\delta_{1,2}=0$ from panel (b) (red arrows), obtained at $\omega_\text{DQD1}=\omega_\text{DQD2}$. (d) Line trace of panel (b) when both DQDs slightly detuned, $\omega_\text{DQD1}\neq\omega_\text{DQD2}$ showing two separate resonances. (e) Qubit spectroscopy of DQD$_1$ ($2t_1/2\pi=5.146$~GHz) \emph{vs.} detuning $\delta_1$. DQD$_2$ is detuned to, $\delta_2/2\pi\gtrsim 20$~GHz. (f) At $\delta_2=0$ qubit spectroscopy \emph{vs.} changing detuning $\delta_1$. (g) Line trace at $\delta_{1,2}=0$ from panel (f) (red arrows) showing one resonance peak. (h) Line trace of panel (f) for detuned DQDs. }
\label{fig:SI:symmetric_g_2J}
\end{figure*}

In Fig.~\ref{fig4:Dispersive_2J_detuning}(a) of the main text, the coherent qubit-qubit exchange interaction as function of the detuning from the resonator $\Delta_\mathrm{r}$ is investigated.
The dark state $\ket{+}_\text{d2}$ remains at fixed transition frequency, as is observed in the experiment indicated by the blue line in Fig.~\ref{fig4:Dispersive_2J_detuning}(a).
In Eq.~\eqref{eq:dark}, the energy $\bar E_{+}$, is constant and it is the bright state $\ket{-}_\text{d2}$ which shifts in energy as function of $\Delta_\mathrm{r}$.

Note that the splitting between eigenstates in the case treated here is different from the standard case treated most commonly in literature,
when assuming the DQDs are resonant and coupled through a second order transition via the resonator.
In that case the diagonal terms in Eq.~\eqref{eq:H2} would be equal, corresponding to a tuning point where the dressed frequencies of both DQDs are resonant.
In that case we find a splitting of $2J = 2g_{1}g_{2}/\Delta_\mathrm{r}$ instead of the value obtained here $(g_{1}^{2}+g_{2}^{2})/\Delta_\mathrm{r}$.
The two cases are equivalent only for equal coupling $g_{1} = g_{2}$.
Indeed, for the data shown in Fig.~\ref{fig4:Dispersive_2J_detuning} we used $2J = 2g_{1}g_{2}/\Delta_\mathrm{r}$
since we have only access to the dispersively shifted frequencies of the DQDs in the measurements.

\section{Complementary data to Fig.~\ref{fig2:OnResonance}} \label{opposite_fig2}

In addition to the data in the main text [Fig.~\ref{fig2:OnResonance}] we show here the vacuum Rabi mode splitting of DQD$_2$ with the resonator, see Fig.~\ref{fig:SI:symmetric_g_resonance}(a) and (c). We found with our master equation fitting that $\omega_\text{r}=\omega_\text{DQD2}$ is realized at $\delta_2/2\pi=0.2$~GHz. This bias point is used to extract the coupling rate from the resonator to the DQD$_2$ $g_2/2\pi=56$~MHz [Fig.~\ref{fig:SI:symmetric_g_resonance}(c)] and is quoted in the main text.
The collective mode coupling, realized by tuning DQD$_1$ ($g_1/2\pi=53$~MHz) into resonance when DQD$_2$ is in vacuum Rabi mode splitting, to obtain the resonant condition $\omega_\mathrm{r}=\omega_\mathrm{DQD2}=\omega_\mathrm{DQD1}$ which is essentially the same bias point as Fig.~\ref{fig2:OnResonance}(c). This indicates that the DQD system is fully tunable via the parameters $2t_{1,2}$ and $\delta_{1,2}$ allowing us to measure data equivalent to that shown in Fig.~\ref{fig2:OnResonance}.

\section{Resonant interaction with \\ unequal coupling rates, $g_1\neq g_2$.} \label{sec:res:uneq}

In the main text in Fig.~\ref{fig2:OnResonance} and Appendix~\ref{opposite_fig2} we present the vacuum Rabi mode splitting and the collective vacuum Rabi mode splitting measurements
when varying the charge detuning of DQD$_1$ and DQD$_2$ for the configuration where both coupling rates are approximately equal, $g_{1,2}/2\pi\approx55$~MHz.
We use a bias point in Fig.~\ref{fig3:Dispersive_2J}(b) and Fig.~\ref{fig4:Dispersive_2J_detuning} where $g_1/2\pi=34$~MHz and $g_2/2\pi=69$~MHz to measure the $2J$ splitting of the virtual photon-mediated qubit-qubit exchange interaction.
The coupling rates used to fit the data in Fig.~\ref{fig3:Dispersive_2J}(b) and Fig.~\ref{fig4:Dispersive_2J_detuning} are obtained from the vacuum Rabi mode splitting measurements shown in Fig.~\ref{fig:SI:opposite_fig2_main}. See Appendix~\ref{sec:para} for a detailed discussion of fitting procedure employed.

In Fig.~\ref{fig:SI:opposite_fig2_main}, we present the on-resonance interaction for this configuration similar to Fig.~\ref{fig2:OnResonance}.
Also here we observe a dark state when $\omega_\mathrm{r}=\omega_\mathrm{DQD1}=\omega_\mathrm{DQD2}$, by tuning the detuning parameters $\delta_{1,2}$. In principle, Figs.~\ref{fig:SI:opposite_fig2_main}(b) and (g) are showing data of very similar experiments as one DQD is in resonance with the resonator and the opposite DQD is tuned in resonance for the data obtained in both panels. The unequal coupling rate to the resonator making the response of the amplitude of the reflection spectrum $|S_{11}|$ quite different. This is visible by the initial small (large) vacuum Rabi splitting at $\delta_{2,(1)}/2\pi=-2$~GHz set by the coupling rate $g_{1,(2)}$. The collective mode coupling to the resonator is $g_\mathrm{c}/2\pi=\sqrt{g_1^2+g_2^2}/2\pi=76$~MHz.

In addition we present line traces in Fig.~\ref{fig:SI:opposite_fig2_main} for the vacuum Rabi splitting [$\omega_\text{r}=\omega_\text{DQDk}$, panel (c) and (h)],
collective mode coupling [$\omega_\mathrm{r}=\omega_\mathrm{DQD1}=\omega_\mathrm{DQD2}$, panel (e) and (j)]
and the case where the $|+\rangle_\text{r2}$-state is approximately resonant with the DQD which is being tuned [$\omega_\mathrm{DQD2}+g_1\approx\omega_\mathrm{DQD2}$ in panel (d) and $\omega_\mathrm{DQD1}+g_2\approx\omega_\mathrm{DQD1}$ in panel (i)].
The later case shows clearly the difference in coupling strengths to the resonator of both DQDs.
This difference is fully captured by our master equation simulations (solid and dashed line in Fig.~\ref{fig:SI:opposite_fig2_main}).

\section{Spectroscopy of dispersive qubit-qubit interaction for the $2g_1\approx g_2$ configuration.}

In Figs.~\ref{fig3:Dispersive_2J}(a) and (b) of the main text, the DQD$_1$ and DQD$_2$ are tuned into resonance, fulfilling the condition, $\omega_\text{DQD1}=\omega_\text{DQD2}$ resulting in the hybridized states, $|\pm\rangle_\text{d2}$. With DQD$_2$ (DQD$_1$) largely detuned by setting $\delta_{1,2}/2\pi\gtrsim20$~GHz, we observe the DQDs single charge qubit behaviour since we can fit it by the expected spectrum, $\omega_\text{DQD}=\sqrt{4t^2+\delta^2}$, see Fig.~\ref{fig:SI:asymmetric_qubit_spec}. This demostrates full gate control of each DQD qubit and excludes coupling to spurious two-level fluctuators~\cite{Lisenfeld2015}.							
In addition we presented the specular tuning of the DQDs compare to the data shown in Fig.~\ref{fig3:Dispersive_2J}(b) and Fig.~\ref{fig:SI:asymmetric_qubit_spec}.
Here, the DQDs are tuned to $2t_1/2\pi=4.453$~GHz and $2t_2/2\pi=4.472$~GHz, $\delta_2=0$ realizing resonance ($\omega_\text{DQD1}=\omega_\text{DQD2}$) at finite detuning ($\delta_1/2\pi=\pm0.4$~GHz) showing clear hybridization between the two qubit state via virtual photon exchange [Fig.~\ref{fig:SI:asymmetric_qubit_spec}(c)]. Here, qubit spectroscopy shows different qubit contrast for the two DQDs, which is attributed to the difference in coupling rate, $g_{1,2}$. At detuning, $|\delta_1/2\pi|=\pm$~0.6~GHz, indicated by the orange dots in Fig.~\ref{fig:SI:asymmetric_qubit_spec}(c), the bare qubit frequencies are equal, resulting in a dark state.
The resonance frequencies, in Fig.~\ref{fig:SI:asymmetric_qubit_spec}(c) are fitted to the full Hamiltonian model for the interacting ($g_{k}\neq 0$ red-dashed line) and non-interacting case ($g_{k}=0$ green-dashed line), showing quantitative agreement with the data.
The fact that the lower red-dashed line in panel (c) does not converge to the lower green-dashed line for large $|\delta_1|$, can be attributed to the breakdown of the dispersive approximation in this regime.

The tunnel rates, $2t_{1,2}$, obtained to the data in Fig.~\ref{fig:SI:asymmetric_qubit_spec}(a), B are not exactly the same as in Fig.~\ref{fig3:Dispersive_2J}(b) and Fig.~\ref{fig:SI:asymmetric_qubit_spec}(c), since this data was measured in a separate run (4~weeks) later in the same bias configuration.	

\

\section{Coherent qubit-qubit exchange interaction in spectroscopy with equal coupling rate, $g_1\approx g_2$.}\label{sec:2J_equal}

In the main text, we used the configuration $g_1/2\pi=34$~MHz and $g_2/2\pi=69$~MHz, to demonstrate the coherent qubit-qubit exchange interaction in qubit spectroscopy. At the resonance condition, $\omega_\mathrm{DQD1}=\omega_\mathrm{DQD2}$ the $|+\rangle_\text{d2}$ state is a dark state for two equally coupled DQDs. We verify this by using this bias point to demonstrate two-qubit interaction in Fig.~\ref{fig3:Dispersive_2J}(a) resulting in a dark state. Here, we present additional data and line traces to support our finding. In Fig.~\ref{fig:SI:symmetric_g_2J}(a) and (e) we show that both DQDs display the typical level structure of a charge qubit with $\omega_\text{DQD}=\sqrt{4t^2+\delta^2}$.

Fixing one DQD at zero detuning $\delta=0$ and tuning the other one we observe virtual photon qubit-qubit exchange interaction resulting in hybridized states. The higher frequency state $|+\rangle_\text{d2}$ is dark since it is anti-symmetric and thus cannot be excited by symmetric probe to both DQDs with the same phase. Correspondingly, on resonance only single resonances are observed in the linetraces in Figs.~\ref{fig:SI:symmetric_g_2J}(c) and (g). By detuning one of the DQDs, the second resonance can be excited as well again, Figs.~\ref{fig:SI:symmetric_g_2J}(d) and (h). The effect is fully captured by our master equation simulations (lines in Fig.~\ref{fig:SI:symmetric_g_2J}). In the main text we instead discuss the device tuned to a bias point where $g_1\neq g_2$, see Fig.~\ref{fig3:Dispersive_2J}(b) and~\ref{fig:SI:asymmetric_qubit_spec}(c).

\section{Description of fitting procedure and extracted parameters.}\label{sec:para}

\begin{table*}
\begin{tabular}{|c@{ }l|rl|rl|rl|rl|rl|rl|}

\multicolumn{2}{c}{}&  \multicolumn{2}{c}{Fig.~\ref{fig:SI:opposite_fig2_main}(a)}&\multicolumn{2}{c}{Fig.~\ref{fig:SI:opposite_fig2_main}(b)}&\multicolumn{2}{c}{Fig.~\ref{fig:SI:opposite_fig2_main}(c), (d), (e)}&  \multicolumn{2}{c}{Fig.~\ref{fig:SI:opposite_fig2_main}(f)}&\multicolumn{2}{c}{Fig.~\ref{fig:SI:opposite_fig2_main}(g)}&\multicolumn{2}{c}{Fig.~\ref{fig:SI:opposite_fig2_main}(h), (i), (j)}   \\ \hline
  $\omega_\text{r}/2\pi$&(MHz)& 4461.8&$\pm$  0.5& 4476&$\pm$~3& 4476&* & 4447.8&$\pm$0.1& 4463&$\pm$~1& 4463&*   \\ \hline
  $\kappa_\text{int}/2\pi$& (MHz)&& - & &-& 8.4&$\pm$~0.2&           & - & &-& 9.1&$\pm$~0.2    \\ \hline
  $\kappa_\text{ext}/2\pi$& (MHz)&& - & &-& 2.64&$\pm$~0.03&         & - & &-& 2.64&$\pm$~0.03 \\ \hline \hline
  $2t_2/2\pi$&(MHz) & 4447.7& $\pm$ 0.5 & 4451&$\pm$~2& 4451&*       & & -  & 4463&$\pm$~2& 4463&*\\ \hline
  $g_2/2\pi$&(MHz) & 69.8 &  $\pm$ 0.4 & 69.0&$\pm$~0.5& 69.0&* &  & -  & 69.3&$\pm$~0.5& 69.3&*    \\ \hline
  $\gamma_{2,2}\cc/2\pi$&(MHz) & & - & &-& 5.5&$\pm$~0.3&            & - & &-& 4.0&$\pm$~0.6 \\ \hline \hline
  $2t_1/2\pi$&(MHz)&&-&4452&$\pm$~1&4452 &*                          &4450.1&$\pm$~0.2&4461&$\pm$~1&4461 &*\\ \hline
  $g_1/2\pi$ &(MHz)&&-&33.2&$\pm$~0.6& 33.2&*                   &  33.67&$\pm$~0.08&34.7&$\pm$~0.4& 34.7&*\\ \hline
  $\gamma_{2,1}\cc/2\pi$&(MHz)&&-&&-& 5.3&$\pm$~0.2&                 & -&&-& 6.9&$\pm$~0.3 \\
  \hline
\end{tabular}
\caption{Extracted values from the data shown in Fig.~\ref{fig:SI:opposite_fig2_main} by fitting the linetraces as described in text. For the parameters indicated by Fig.~\ref{fig:SI:opposite_fig2_main}(a) and (b) a Hamiltonian fit to the resonance positions was performed. For the parameter indicated with Fig.~\ref{fig:SI:opposite_fig2_main}(c), (d), (e) a master equation fit was preformed by fixing the parameters obtained from the previous fit (indicated by *). For the parameters indicated by Fig.~\ref{fig:SI:opposite_fig2_main}(f) and (g) a Hamiltonian fit to the resonance positions was performed. For the parameter indicated with Fig.~\ref{fig:SI:opposite_fig2_main}(c), (d) and (e) a master equation fit was preformed by fixing the parameters obtained from the previous fit (indicated by *).}
\label{tab:para_figS}
\end{table*}

\begin{table}
\begin{tabular}{|c@{ }l|rl|rl|rl|}

\multicolumn{2}{c}{}&  \multicolumn{2}{c}{Fig.~\ref{fig2:OnResonance}(a), (b)}&\multicolumn{2}{c}{Fig.~\ref{fig2:OnResonance}(d)}&\multicolumn{2}{c}{Fig.~\ref{fig2:OnResonance}(c)}  \\ \hline
  $\omega_\text{r}/2\pi$&(MHz)& 5170&$\pm$  1& 5172&$\pm$~1& 5172&*   \\ \hline
  $\kappa_\text{int}/2\pi$& (MHz)& 18& $\pm$ 2 & 17&$\pm$~1& 17&*   \\ \hline
  $\kappa_\text{ext}/2\pi$& (MHz)& 6.5& $\pm$ 0.1 & 6.1&$\pm$~0.1& 6.1&* \\ \hline \hline
  $2t_1/2\pi$&(MHz) & 5166& $\pm$ 1 & 5138&$\pm$~1& 5138&*\\ \hline
  $g_1/2\pi$&(MHz) & 53.4 &  $\pm$ 0.2 & 51.1&$\pm$~0.4& 51.1&*  \\ \hline
  $\gamma_{2,1}\cc/2\pi$&(MHz) & 5.3& $\pm$ 0.9 & 6.4&$\pm$~1.2& 6.4&* \\ \hline \hline
  $2t_2/2\pi$&(MHz)&&-&&-& 5156.2&$\pm$~0.6\\ \hline
  $g_2/2\pi$ &(MHz)&&-&&-& 56.7&$\pm$~0.2\\ \hline
  $\gamma_{2,2}\cc/2\pi$&(MHz)&&-&&-& 6.0&$\pm$~0.6 \\
  \hline
\end{tabular}
\caption{Extracted values from the data shown in Fig.~\ref{fig2:OnResonance} by fitting the linetraces as described in text. For Fig.~\ref{fig2:OnResonance}(a), (b) the line trace at $\delta_1=0$ was used. We extract the parameters of the data presented in Fig.~\ref{fig2:OnResonance}(d) at detuning $\delta_2/2\pi=-2.9$~GHz. The extracted parameters in column Fig.~\ref{fig2:OnResonance}(c) are those of DQD$_2$ with resonator and DQD$_1$ parameters fixed (indicated by *).}
\label{tab:para_fig2}
\end{table}
\begin{table}[!t]
\begin{tabular}{|c@{ }l|rl|rl|rl|}

\multicolumn{2}{c}{}&  \multicolumn{2}{c}{Fig.~\ref{fig:SI:symmetric_g_resonance}(a), (c)}&\multicolumn{2}{c}{Fig.~\ref{fig:SI:symmetric_g_resonance}(b)}&\multicolumn{2}{c}{Fig.~\ref{fig:SI:symmetric_g_resonance}(e)}  \\ \hline
  $\omega_\text{r}/2\pi$&(MHz)& 5170.9&$\pm$  0.8& 5167.8&$\pm$~0.9& 5167.8&*   \\ \hline
  $\kappa_\text{int}/2\pi$& (MHz)& 12& $\pm$ 2 & 10&$\pm$~2& 10&*   \\ \hline
  $\kappa_\text{ext}/2\pi$& (MHz)& 5.7& $\pm$ 0.1 & 5.7&$\pm$~0.1& 5.7&* \\ \hline \hline
  $2t_2/2\pi$&(MHz) & 5167.9& $\pm$ 0.8 & 5157.3&$\pm$~0.8& 5157.3&*\\ \hline
  $g_2/2\pi$&(MHz) & 55.9 &  $\pm$ 0.3 & 55.8&$\pm$~0.3& 55.8&*  \\ \hline
  $\gamma_{2,2}\cc/2\pi$&(MHz) & 11& $\pm$ 2 & 12&$\pm$~2& 12&* \\ \hline \hline
  $2t_1/2\pi$&(MHz)&&-&&-& 5183.1&$\pm$~0.7\\ \hline
  $g_1/2\pi$ &(MHz)&&-&&-& 53.4&$\pm$~0.3\\ \hline
  $\gamma_{2,1}\cc/2\pi$&(MHz)&&-&&-& 5.7&$\pm$~0.6 \\
  \hline
\end{tabular}
\caption{Extracted values from the data shown in Fig.~\ref{fig:SI:symmetric_g_resonance} by fitting the linetraces as described in text. For Fig.~\ref{fig:SI:symmetric_g_resonance}(a), (c) the line trace at $\delta_2/2\pi=0.2$~GHz was used. For extracting the values indicated as Fig.~\ref{fig:SI:symmetric_g_resonance}(b) we use the data in this figure at detuning $\delta_1/2\pi=-2.9$~GHz, most left line trace. The extracted parameters indicated as Fig.~\ref{fig:SI:symmetric_g_resonance}(e) were use to extract the DQD$_1$ parameters with resonator and DQD$_2$ parameters fixed (indicated by *).}
\label{tab:para_figS2}
\end{table}

\begin{table}[h!]
\begin{tabular}{|c@{ }l|rl|rl|}

  \multicolumn{2}{c}{}&  \multicolumn{2}{c}{Fig.~\ref{fig3:Dispersive_2J}(a)}&\multicolumn{2}{c}{Fig.~\ref{fig3:Dispersive_2J}(b)} \\ \hline
  $\omega_\text{r}/2\pi$ &(MHz)& 5432.9&$\pm$~0.9 &4650&$\pm$~3 \\ \hline
  $2t_1/2\pi$ &(MHz)& 5168.3&$\pm$~0.4 & 4461&$\pm$~1\\ \hline
  $2t_2/2\pi$ &(MHz) &5164.1&$\pm$~0.3&4440&$\pm$~1 \\
  \hline
\end{tabular}
\caption{Extracted values from the fits performed on the data presented in Fig.~\ref{fig3:Dispersive_2J}(a) with taking the coupling rates obtained from the fits in Fig.~\ref{fig2:OnResonance}. For Fig.~\ref{fig3:Dispersive_2J}(b) the coupling rates extracted from the data in Fig.~\ref{fig:SI:opposite_fig2_main}. In the caption Fig.~\ref{fig3:Dispersive_2J} we quote the Lamb shifted qubit transition as they are here in the table directly extracted from the Hamiltonian fit.}
\label{tab:para_fig3A}
\end{table}

\begin{table}[h!]
\begin{tabular}{|c@{ }l|rl|rl|rl|}

  \multicolumn{2}{c}{}&  \multicolumn{2}{c}{Fig.~\ref{fig:SI:asymmetric_qubit_spec}(a)}&\multicolumn{2}{c}{Fig.~\ref{fig:SI:asymmetric_qubit_spec}(b)}&\multicolumn{2}{c}{Fig.~\ref{fig:SI:asymmetric_qubit_spec}(c)} \\ \hline
  $\omega_\text{r}/2\pi$ &(MHz)& 4713&$\pm$~2 &4697&$\pm$~1&4670&$\pm$~2 \\ \hline
  $2t_1/2\pi$ &(MHz)& 4392.2&$\pm$~0.3 &&-& 4493.1&$\pm$~0.3\\ \hline
  $2t_2/2\pi$ &(MHz) &&-&4468.7&$\pm$~0.4&4456.3&$\pm$~0.3 \\
  \hline
\end{tabular}
\caption{Parameters extracted from the fits performed in Fig.~\ref{fig:SI:asymmetric_qubit_spec} with coupling rated obtained from fits of the data in Fig.~\ref{fig2:OnResonance}.}
\label{tab:para_figS5}
\end{table}

\begin{table*}[!t]
\begin{tabular}{|c@{ }l|rl|rl|rl|rl|rl|rl|}

  \multicolumn{2}{c}{}&  \multicolumn{2}{c}{Fig.~\ref{fig:SI:symmetric_g_2J}(a)}&\multicolumn{2}{c}{Fig.~\ref{fig:SI:symmetric_g_2J}(b)}&  \multicolumn{2}{c}{Fig.~\ref{fig:SI:symmetric_g_2J}(c), (d)}&\multicolumn{2}{c}{Fig.~\ref{fig:SI:symmetric_g_2J}(e)}&  \multicolumn{2}{c}{Fig.~\ref{fig:SI:symmetric_g_2J}(f)}&\multicolumn{2}{c}{Fig.~\ref{fig:SI:symmetric_g_2J}(g), (h)}  \\ \hline
  $\omega_\text{r}/2\pi$ &(MHz)& 5443.7&$\pm$~0.7 &5432.9&$\pm$~0.9& 5432.9&* &5443&$\pm$~2&5432&$\pm$~1&5432&*  \\ \hline
  $2t_1/2\pi$ &(MHz)& 5145.8&$\pm$~0.2 & 5168.3&$\pm$~0.4&5168.6&$\pm$~0.4 & 5168.8&$\pm$~0.7&5162.2&$\pm$~0.3&5160.0&$\pm$~0.4\\ \hline
  $\gamma_{2,1}/2\pi$& (MHz)&&-&&-&12.0&$\pm$~0.5&&-&&-&13.7&$\pm$~0.6 \\ \hline
  $2t_2/2\pi$ &(MHz) &&-&5162.8&$\pm$~0.7&5164.1&$\pm$~0.3 & &-&5159.7&$\pm$~0.5&5156.0&$\pm$~0.5 \\ \hline
  $\gamma_{2,2}/2\pi$& (MHz)&&-&&-&8.8&$\pm$~0.5&&-&&-&9.6&$\pm$~0.7 \\
  \hline
\end{tabular}
\caption{Extracted values from the Hamiltonian [Fig.~\ref{fig:SI:symmetric_g_2J} panel (a), (b), (e) and (f)] and master equation fits [Fig.~\ref{fig:SI:symmetric_g_2J} panel (c), (d), (g) and (h)]. Parameters keeping fixed are indicated by *.}
\label{tab:para_figS6}
\end{table*}

In this section we describe the procedure we used for fitting the data and extracting the master equation parameters.

In general, if not stated otherwise, fits are least-square fits to the full master equation input-output model of the system, Eq.~\eqref{eq:MESLH}.
In order to reduce the number of independent fit parameters, we adopt an iterative approach, where we successively extract different fit parameters from different parts of the spectrum.

We illustrate the procedure using the example of Fig.~\ref{fig2:OnResonance} (Tab.~\ref{tab:para_fig2}). The same procedure was followed for the data shown in Fig.~\ref{fig:SI:symmetric_g_resonance} (Tab.~\ref{tab:para_figS2}). Initially, we use the leftmost line trace of Fig.~\ref{fig2:OnResonance}(a), which is a resonator-like resonance, to obtain initial estimates for the resonator decay rates $\kappa_\text{int}$ and $\kappa_\text{ext}$ as well as the resonator frequency $\omega_\text{r}$.
These values are then used as initial parameters to fit the line trace shown in Fig.~\ref{fig2:OnResonance}(b), close to resonance between DQD$_1$ and the resonator, $\omega_\text{r}=\omega_\text{DQD1}$. The quoted frequencies in the main text are the measured Lamb shifted frequencies~\cite{Blais2004}.
From this fit we extract the DQD coupling strength $g_{1}$ and its linewidth $\gamma_{2,1}\cc = \gamma_{1,1}/2 + \gamma_{2,1}$, see Tab.~\ref{tab:para_fig2}. Note that in all the fits we have set the pure dephasing for each DQD to zero, $\gamma_{2,k}=0$, as its effect on the scattering spectrum is essentially indistinguishable from the relaxation rates $\gamma_{1,k}$. The essence of the fits is to capture the linewidth of the resonances, given by $\gamma_{2,1}\cc$. In principle, taking into account the linetraces at finite detuning ($\delta\neq0$), could additionally provide insight into the relaxation and dephasing rates.
As here this was not essential to obtain more accurate fits, we decided to keep $\gamma_{2,k}$ fixed for simplicity.
The extracted DQD linewidth $\gamma_{2,k}$ is measured at finite power and is close to the extrapolated zero power limit $\Gamma_{2,1}/2\pi=4.8\pm0.6$~MHz, measured independently in the dispersive regime (see main text).
We generally observe that the extracted qubit linewidths are powerbroadend as the values are typically 1-3~MHz higher than $\Gamma_{2,1}/2\pi$ compare for example with the values in Tab.~\ref{tab:para_fig2}.

We also extract the resonator internal and external loss rates from independent measurements (not shown) with both DQDs detuned ($\delta_{1,2}/2\pi\gtrsim20$~GHz),
and find $\kappa_\text{int}/2\pi=18.7\pm0.5$~MHz and $\kappa_\text{ext}/2\pi=7.4\pm0.2$~MHz for $\omega_\text{r}/2\pi=5.170$~GHz which is comparable to what is obtained from the data in Fig.~\ref{fig2:OnResonance} and listed in Tab.~\ref{tab:para_fig2}.
Finally, to calibrate the detuning axis, we perform a simultaneous fit to three different line traces (not shown) of Fig.~\ref{fig2:OnResonance}(a), using the parameters from the previous fit.

Even though the gate settings are exactly the same for DQD$_1$ in the measurements shown in Figs.~\ref{fig2:OnResonance}(b) and ~\ref{fig2:OnResonance}(d), as the later one was measured 2 days later a small frequency shift (30~MHz) of the DQD tunnel rate was extracted from the fit. We attribute this shift to changes in the environmental offset charge distribution, influencing the effective applied gate voltages, which effectively shift the tunnel rate $2t_{1}$. The shift is small but has to be taken into account to improve the quality of the fits, compare the values in Tab.~\ref{tab:para_fig2}.
Fitting to the data in Fig.~\ref{fig2:OnResonance}(d), we start with a single line trace at large negative detuning ($\delta_2/2\pi=-2.9$~GHz), where the resonator and DQD$_1$ are resonant. We fit the resonator and DQD$_1$ parameters to this line trace and use those parameters as fixed (indicated by * in Tab.~\ref{tab:para_fig2}) when obtaining the parameters for DQD$_2$ from a fit at $\delta_2=0$, where all three systems are close to resonant. The extracted parameters are displayed in Tab.~\ref{tab:para_fig2}.

To obtain the parameter values for the data presented in Fig.~\ref{fig:SI:opposite_fig2_main}, the procedure was modified slightly.
We first extracted resonance positions from the experimental data using a simple fit to Lorentzians, and then used a pure Hamiltonian model, Eq.~\eqref{eq:HTot}, to fit all the Hamiltonian parameters to the spectrum.
Then we applied the master equation simulation with these parameters as input to single line traces of the data to obtain the linewidths. All parameters extracted from the data presented in Fig.~\ref{fig:SI:opposite_fig2_main} are displayed in Tab.~\ref{tab:para_figS}.

With this procedure, agreement between theory and experiment is not quite as good as for the full master equation simulations used for Figs.~\ref{fig2:OnResonance} and~\ref{fig:SI:symmetric_g_resonance}, but it needs far less computational effort. The main difference between the two methods is that the master  equation simulation is more sensitive to residual detuning effects due to non-zero $\delta_{k}$ parameters then the Hamiltonian fitting, since it can also take into account the amplitude correctly, leading to an overall better fit.

Finally, for the data presented in Fig.~\ref{fig3:Dispersive_2J} we extract frequencies of resonances in the experimental data using a simple Lorentzian fit.
We then use the coupling parameters $g_{1,2}$ obtained from spectroscopy of the system configuration, shown in Fig.~\ref{fig:SI:opposite_fig2_main} and fit this data to a Hamiltonian model, Eq.~\eqref{eq:HTot}, where the only free parameters are the resonator and qubit frequencies as well as the scale of the detuning axis. Results are shown in Tab.~\ref{tab:para_fig3A}. The same fitting procedure was used for the extracted parameters presented in Tabs.~\ref{tab:para_figS5} and~\ref{tab:para_figS6}


For the fits to the virtual photon-mediated coupling between the two DQDs in the dispersive regime [Fig.~\ref{fig4:Dispersive_2J_detuning}] we again start by extracting resonance frequencies from the data for both the DQD-like resonances as well as the detuned resonator. Using the DQD parameters extracted from the data in Fig.~\ref{fig3:Dispersive_2J} as input, we then fit each of these datapoints to a Hamiltonian model to extract the DQD tunnel rates $2t_{k}$ (assuming $\delta_k = 0$) and resonator frequency $\nu_{\text{r}}$.
Finally we fit the $2J$ data to a linear dependence in $1/\Delta_\text{r}$, shown in Fig.~\ref{fig4:Dispersive_2J_detuning}(a). We observe linear dependence with slight departure from the data.\\


\textbf{Acknowledgements:} We acknowledge contributions by C.~K. Andersen, S. Gasparinetti, M. Collodo, J. Heinsoo, S. Storz, M. Frey, A. Stockklauser and M. Gabureac. We thank Ata\c{c} \.{I}mamo\u{g}lu and Gianni Blatter for valuable feedback on the manuscript.\\

\textbf{Funding:} This work was supported by the Swiss National Science Foundation (SNF) through the National Center of Competence in Research (NCCR) Quantum Science and Technology (QSIT), the project Elements for Quantum Information Processing with Semiconductor/Superconductor Hybrids (EQUIPS) and by ETH Zurich. \\

\textbf{Authors contributions:} PS, JHU and JVK designed the device with input from AW. DJvW, PS and JVK fabricated the device. Electrical and microwave measurements and data analysis were performed by DJvW, PS and JHU. The theoretical model and fits were done by CM. AJL, TI and KE had valuable input to the experiments. CR and WW grew the GaAs heterostructure. The manuscript was written by DJvW, CM and AW with comments from all authors. AW supervised the project.\\

\textbf{Data and materials availability:} The data presented in this paper and corresponding supplementary material will be available online at ETH Zurich repository for research data, https://www.research-collection.ethz.ch/.

\bibliographystyle{naturemag}
\bibliography{ReferenceDatabase}

\end{document}